\documentclass[aps,prd,nofootinbib,eqsecnum,10pt]{revtex4-2}

\usepackage{graphicx}
\usepackage{bm}
\usepackage{amsfonts}
\usepackage{amsmath,amsthm,amssymb}
\usepackage{xcolor}
\usepackage{dcolumn}
\usepackage{nameref}
\usepackage{ulem}
\usepackage{hyperref}
\hypersetup{
  setpagesize=false,
  bookmarksnumbered=true,%
  bookmarksopen=true,%
  colorlinks=true,%
  linkcolor=blue,
  linktoc=all,
  citecolor=red,
}
\allowdisplaybreaks[1]

\begin{document}

\newcommand{\newc}{\newcommand}

\newc{\be}{\begin{equation}}
\newc{\ee}{\end{equation}}
\newc{\ba}{\begin{eqnarray}}
\newc{\ea}{\end{eqnarray}}
\newc{\D}{\partial}
\newc{\Mpl}{M_{\rm pl}}
\newc{\tF}{\tilde{F}}
\newc{\da}{\delta A}
\newc{\rk}[1]{\textcolor{red}{#1}}
\newc{\kt}[1]{\textcolor{teal}{#1}} 
\newc{\sn}[1]{\textcolor{brown}{#1}}
\newc{\cosq}{c_{\Omega}^2}

\title{
Linear perturbations of dyonic black holes
\texorpdfstring{\\}{} 
in the lowest-order \texorpdfstring{$U(1)$}{U(1)} gauge-invariant scalar-vector-tensor theories}

\author{Kitaro Taniguchi, Shunta Nishimura, Naoki Tsukamoto, and Ryotaro Kase}

\affiliation{Department of Physics, Faculty of Science, 
Tokyo University of Science, 1-3, Kagurazaka,
Shinjuku-ku, Tokyo 162-8601, Japan}

\date{December 11, 2025}

\begin{abstract}
We study linear perturbations on top of the static and spherically symmetric background of dyonic black hole solutions endowed with electric and magnetic charges, as well as a scalar hair, in the lowest-order $U(1)$ gauge-invariant scalar-vector-tensor theories. 
The presence of magnetic charges in the background solutions gives rise to a mixing between the odd-parity and even-parity sectors of perturbations, which makes it impossible to analyze each sector separately. 
Thus, we expand the action up to second order in both odd-parity and even-parity perturbations and derive the general conditions for the absence of ghosts and Laplacian instabilities.
We apply these general conditions to extended Einstein-Maxwell-scalar theories, which encompass numerous types of concrete models from the literature known to have dyonic black hole solutions with the scalar hair, and examine their stabilities. 
Our general framework for studying stability conditions and dynamics of perturbations can be applied to a wide variety of theories, including nonlinear electrodynamics coupled to a scalar field, as well as to calculations of black hole quasinormal modes.
\end{abstract}


\maketitle

\section{Introduction}
\label{introsec}

General relativity (GR) describes gravity in solar-system scales with a high precision \cite{Will:2014kxa} 
while the validity of GR in nonlinear regimes of gravity has not yet been investigated enough. 
In recent years, observations related to the strong field of gravity have developed rapidly.
For examples, the LIGO-Virgo-KAGRA Collaborations enable us to detect gravitational waves (GWs) from compact binary coalescences directly \cite{LIGOScientific:2016aoc, LIGOScientific:2016sjg, LIGOScientific:2017bnn, LIGOScientific:2017vox, LIGOScientific:2017ycc, LIGOScientific:2017vwq, LIGOScientific:2018mvr, LIGOScientific:2020stg, LIGOScientific:2020aai, LIGOScientific:2020zkf, LIGOScientific:2021qlt, KAGRA:2021vkt}, and the Event Horizon Telescope captured the shadow images of black hole (BH) candidates in centers of galaxies \cite{EventHorizonTelescope:2019dse, EventHorizonTelescope:2022wkp}.
These observations make progress in the exploration of physics associated with BHs, 
and GR is consistent with them so far.
On the other hand, in the context of cosmology, the finding of the late-time cosmic acceleration motivates one to introduce a new degree of freedom in gravitational theories \cite{Peebles:2002gy, Copeland:2006wr, Sotiriou:2008rp, DeFelice:2010aj, Clifton:2011jh,Heisenberg:2018vsk,Kase:2018aps}.
\par
In a static and spherically symmetric (SSS) spacetime, for a minimally coupled canonical scalar field \cite{Hawking:1971vc, Bekenstein:1972ny} and k-essence \cite{Graham:2014mda} as well as for a nonminimally coupled scalar field with the Ricci scalar $R$ of the form $F (\phi) R$ \cite{Hawking:1972qk, Bekenstein:1995un, Sotiriou:2011dz, Faraoni:2017ock}, it is known that the scalar field profile settles down to be trivial. 
Obtaining nontrivial configurations of the scalar field in black hole solution is not a simple problem \cite{Hui:2012qt, Babichev:2016rlq}.
On the other hand, this is not the case in the presence of the electromagnetic field 
which can admit nontrivial configurations of a vector field in the black hole spacetime.
Requiring a stationary and asymptotically flat ansatz in the Einstein-Maxwell system, 
the resultant black hole spacetime is characterized by only three parameters: mass $M$, 
electric (or magnetic) charge $Q$, and angular momentum $L$ \cite{Israel:1967wq, Carter:1971zc, Ruffini:1971bza, Hawking:1971vc}. 
This statement, called the no-hair theorem, does not necessarily hold in the presence of 
a scalar field coupled to a vector field. 
For instance, the dilaton field arising from the compactification of heterotic string theories universally couples to other fields including the Maxwell field, and 
yields a hairy solution, the so-called Gibbons-Maeda-Garfinkle-Horowitz-Strominger (GMGHS) solution, in a static and spherically symmetric spacetime \cite{Gibbons:1982ih, Gibbons:1987ps, Garfinkle:1990qj}.
The hairy BH solutions under the existence of direct scalar-vector couplings have been broadly studied in the literature~\cite{Gibbons:1982ih, Gibbons:1987ps, Garfinkle:1990qj, Campbell:1991rz, Lee:1991jw, Reuter:1991cb, Monni:1995vu, Stefanov:2007qw, Stefanov:2007bn, Stefanov:2007eq, Sheykhi:2014gia, Sheykhi:2014ipa, Sheykhi:2015ira, Fan:2015oca, Dehghani:2018eps, Herdeiro:2018wub, Myung:2018vug, Herdeiro:2019oqp, Myung:2019oua, Konoplya:2019goy, Hod:2019ulh, Dehghani:2019cuf, Filippini:2019cqk, Boskovic:2018lkj, Fernandes:2019kmh, Astefanesei:2019pfq, Herdeiro:2019tmb, Brihaye:2019gla, Fernandes:2020gay, Mahapatra:2020wym, Priyadarshinee:2021rch, Priyadarshinee:2023cmi, Promsiri:2023yda, Belkhadria:2023ooc, Bakopoulos:2024hah, Taniguchi:2024ear}.
\par
Introducing a scalar field as well as vector and tensor fields and demanding the equations of motion to be up to second order for derivatives in a general curved spacetime, 
Heisenberg~\cite{Heisenberg:2018acv} derived the general action of scalar-vector-tensor (SVT) theories which can describe direct interactions between scalar and vector fields in a unified way.
The SVT theories are classified into two classes according to the presence or absence of the $U(1)$ gauge invariance. 
If the vector field respects the $U(1)$ gauge symmetry, the subclass is called the $U(1)$ gauge-invariant (GI) SVT theories, to which the aforementioned scalar-vector couplings that give rise to hairy black hole solutions belong.
In the scope of $U(1)$ GI SVT theories, the authors of Ref.~\cite{Heisenberg:2018vti} showed the existence of hairy BH solutions possessing scalar and electric charges. 
On that extension, the authors in Refs.~\cite{Heisenberg:2018mgr, Zhang:2024cbw} studied the linear perturbations on top of the SSS background, and performed a stability analysis for these hairy solutions.
The Einstein-Maxwell theories allow a Reissner-Nordstr\"om solution to possess not only a electric charge but also a magnetic charge.
Magnetic monopoles \cite{Dirac:1931kp} often appear in unified gauge theories \cite{tHooft:1974kcl, Polyakov:1974ek} and string theories \cite{Wen:1985qj}.
In Ref.~\cite{Maldacena:2020skw}, it is pointed out that extremal magnetic BHs are more stable than their electric counterparts.
The BHs possessing electric and magnetic charges, the so-called dyonic BHs, were well studied in the context of string theories \cite{Gibbons:1987ps}.
Following those works, it is of interest to explore dyonic BH solutions in $U(1)$ GI SVT theories.
In Ref.~\cite{Taniguchi:2024ear}, a new type of dyonic BH solutions with a scalar hair was found in a subclass of the lowest-order $U(1)$ GI SVT theories.
The dilatonic coupling which produces the GMGHS solution can be included in the general theories.
The authors of Refs.~\cite{Kase:2023kvq, Pope:2024ncb} showed the linear stability of electrically charged GMGHS solution.
The magnetically charged one is also studied in terms of gravitational lensing \cite{Bhadra:2003zs}, accretion disks \cite{Karimov:2018whx}, and particle orbits \cite{Lungu:2024ewz, Tsukamoto:2024asy}.
However, the stability of the magnetic GMGHS solution itself has not been confirmed.
It is worthwhile to construct a framework for studying the dynamics of linear perturbations 
on top of the SSS background, based on the general lowest-order $U(1)$ GI SVT theories in the presence of both electric and magnetic charges, which enables one to examine the stability of a wide variety of dyonic hairy black hole solutions studied in the literature so far in a unified manner.
\par
In this paper, we study black hole perturbations in the lowest-order $U(1)$ GI SVT theories with a vector field which is charged both electrically and magnetically.
The linear perturbations on top of the SSS spacetime can be expanded in the spherical harmonics \cite{Regge:1957td, Zerilli:1970se, Moncrief:1974gw}.
According to the symmetry against parity transformation, the perturbative variables are 
classified into odd-parity and even-parity types. 
In the absence of the magnetic charge, these two parity sectors decouple from 
each other~\cite{Zerilli:1974ai}, which enables one to analyze the dynamics of 
the odd- and even-parity sectors separately. 
In the presence of a magnetic charge, however, a mixing of the two sectors occurs 
in the second-order action for perturbations~\cite{Kodama:2003kk, Nomura:2020tpc, Pereniguez:2023wxf, Chen:2024hkm}.
Due to this mixing, the dynamics of the odd- and even-parity sectors must be analyzed simultaneously in order to study the stability of dyonic hairy black hole solutions
\footnote{We used “Maple” and “Mathematica” to calculate the expanded second-order action and to derive the stability conditions. The derivation of the stability conditions inevitably involves theory-dependent analytical steps which make the implementation of an automated and unified calculation procedure difficult. Since all the commands we executed are based on fundamental operations, the procedure can be fully reproduced through straightforward hand calculations, in principle, by following the process described in this paper. The core results of our work thus lie in the analytical methods and the derived stability conditions.}.
\par
This paper is organized as follows.
In Sec.~\ref{backsec}, we define the lowest-order $U(1)$ GI SVT theories, and obtain the background equations of motion on the SSS spacetime.
In Sec.~\ref{pertsec}, we expand all the perturbations in terms of the spherical harmonics and eliminate several variables on the use of gauge degrees of freedom (DOFs). We then derive the second-order action for perturbations which is decomposed into three parts of Lagrangians: odd-parity sector, even-parity sector, and the mixing of both sectors. 
In Sec.~\ref{stabilitysec}, we derive the conditions for the absence of ghosts and Laplacian instabilities, eliminating the nondynamical variables.
This section is composed by three parts according to the modes of spherical harmonics.
In Sec.~\ref{modelsec}, applying our general stability conditions to the extended Einstein-Maxwell-scalar theories, 
we discuss the stability of dyonic BH solutions with a scalar hair for the concrete models including those found in Ref.~\cite{Taniguchi:2024ear}.
Sec.~\ref{concludesec} is devoted to conclusions.

\section{the lowest-order \texorpdfstring{$U(1)$}{U(1)} gauge-invariant scalar-vector-tensor theories with electric and magnetic charges}
\label{backsec}

We consider the lowest-order $U(1)$ gauge-invariant scalar-vector-tensor theories 
described by the Einstein-Hilbert action with the lowest-order SVT interacting term $f_2$~\cite{Heisenberg:2018acv} as
\be
S = \int {\rm d}^4 x \sqrt{-g} 
\left[ \frac{\Mpl^2}{2} R + f_2 (\phi, X, F, \tilde{F}, Y)\right]\,,
\label{act}
\ee
where $g$ is a determinant of the metric tensor $g_{\mu\nu}$, 
$\Mpl$ is the reduced Planck mass related to Newton's gravitational 
constant $G$ as $\Mpl=(8\pi G)^{-1/2}$, $R$ is the Ricci scalar. 
The function $f_2$ depends on the scalar field $\phi$ and 
\be
X = - \frac12 \nabla_{\mu} \phi \nabla^{\mu} \phi\,, \qquad
F = - \frac14 F_{\mu \nu} F^{\mu \nu}\,, \qquad
\tF = - \frac14 F_{\mu \nu} \tilde{F}^{\mu \nu}\,,\qquad
Y = \nabla_{\mu} \phi \nabla^{\nu} \phi F^{\mu \alpha} F_{\nu \alpha}\,.
\label{defX}
\ee
Here, $X$ is the kinetic term of the scalar field $\phi$ with a covariant derivative 
operator $\nabla_{\mu}$, $F$ is the inner product of the field strength tensor $F_{\mu\nu}$ 
of the $U(1)$ gauge-invariant vector field while $\tF$ is the inner product of $F_{\mu\nu}$ 
and its dual $\tilde{F}^{\mu \nu}$ defined by 
\be
F_{\mu\nu} = \nabla_{\mu} A_{\nu} - \nabla_{\nu} A_{\mu}\,, \qquad
\tilde{F}^{\mu \nu} 
= \frac12 \mathcal{E}^{\mu\nu\alpha\beta} F_{\alpha\beta}\,,
\ee
where $A_{\mu}$ is the $U(1)$ gauge-invariant vector field and $\mathcal{E}^{\mu\nu\alpha\beta}$ 
is the anti-symmetric Levi-Civita tensor satisfying the normalization 
$\mathcal{E}^{\mu \nu \alpha \beta} \mathcal{E}_{\mu \nu \alpha \beta} = -4!$.

We focus on the static and spherically symmetric background given by the line element, 
\be
{\rm d}s^2 = -f(r) {\rm d}t^2 + h^{-1}(r) {\rm d}r^2 + r^2 ({\rm d}\theta^2 + \sin^2 \theta {\rm d}\varphi^2)\,, 
\label{line}
\ee
where $t$, $r$, and $(\theta,\varphi)$ correspond to time, radial, and angular coordinates, 
respectively, and $f$ and $h$ are functions of $r$. 
According to the underlying symmetry of the spacetime, we assume that the scalar field depends only on the radial coordinate at the level of background, i.e., 
\be
\phi=\overline{\phi}(r)\,,
\label{phiBG}
\ee
where the overbar represents the value on the background. 
On the other hand, we consider the background configuration of the vector field 
\be
\overline{A}_{\mu}=(A_0(r),0,0,-P\cos\theta)\,,
\label{ABG}
\ee
where $A_0$ is the function depending only on the radial coordinate, and $P$ 
corresponds to the magnetic charge. We note that the presence of the magnetic charge 
does not contradict with the underlying symmetry of the background spacetime. 
Indeed, from Eqs.~\eqref{line}-\eqref{ABG}, the quantities $X$, $F$, $\tF$, $Y$, 
defined in Eq.~\eqref{defX}, are evaluated on the background as 
\be
\overline{X} = - \frac{h \phi'^2}{2}\,,\qquad
\overline{F} = \frac{h A_0'^2}{2f} - \frac{P^2}{2r^4}\,,\qquad
\overline{\tF} = -\frac{P A_0'}{r^2} \sqrt{\frac{h}{f}}\,,\qquad
\overline{Y} = - \frac{h^2 \phi'^2 A_0'^2}{f}\,,
\label{XBG}
\ee
where the  prime represents the derivative with respect to $r$. 
Equation~\eqref{XBG} shows that the scalar-vector interaction term $f_2$ in Eq.~\eqref{act} follows 
the static and spherically symmetric configuration of the background spacetime 
even in the presence of the term $-P\cos\theta$ in Eq.~\eqref{ABG}. 
We omit the overbar and employ the notations $f_{2,x}=\D f_2/\D x$, 
$f_{2,xy}=\D^2 f_2/\D x\D y$, $f_{2,xyz}=\D^3 f_2/\D x\D y\D z$ for 
$\{x,y,z\}=\{\phi,X,F,\tF,Y\}$ in the following discussion for the sake of simplicity. 

Varying the action \eqref{act} with respect to $f$, $h$, $\phi$, and $A_0$, 
we obtain the background equations of motion, 
\ba
&&
{\cal E}_{00}\equiv
\left(\frac{h'}{r}+\frac{h -1}{r^2}\right) \Mpl^2
-f_2+\frac{h A_0'^2 f_{2,F}}{f}-\frac{\sqrt{h} P A_0'f_{2,\tF}}{\sqrt{f} r^2}
-\frac{2 h^2 A_0'^2 \phi'^2 f_{2,Y}}{f}=0\,,
\label{back1}
\\
&&
{\cal E}_{11}\equiv
\left(\frac{h f'}{r f}+\frac{h -1}{r^2}\right) \Mpl^2
-f_2-\left(h f_{2,X}+\frac{4 h^2 A_0'^2 f_{2,Y}}{f}\right) \phi'^2
+\frac{h A_0'^2 {f_{2,F}}}{f}-\frac{\sqrt{h} P A_0' f_{2,\tF}}{\sqrt{f} r^2}=0\,,
\label{back2}
\\
&&
{\cal E}_{\phi}\equiv
{\cal J}_{\phi}'-{\cal P}_{\phi}=0\,,
\label{back3}
\\
&&
{\cal E}_{A}\equiv
{\cal J}_{A}'=0\,,
\label{back4}
\ea
respectively, with
\be
{\cal J}_{\phi}=
-\sqrt{\frac{h}{f}}\,r^2\phi'
\left(f f_{2,X}+2 h A_0'^2 f_{2,Y}\right)
\,,\quad
{\cal P}_{\phi}=
\sqrt{\frac{f}{h}}r^2f_{2,\phi}
\,,\quad
{\cal J}_{A}=
\sqrt{\frac{h}{f}}\,r^2A_0'\left(
f_{2,F}-2 h \phi'^{2} f_{2,Y}\right)
-P f_{2,{\tF}}\,.\label{defJp}
\ee
Equation~\eqref{back4} shows that the vector field current ${\cal J}_A$ is conserved 
by virtue of the $U(1)$ gauge symmetry. On the other hand, the $\phi$ dependence 
in $f_2$ causes the lack of the shift symmetry $\phi\to\phi+\phi_0$, 
where $\phi_0$ is a constant, which makes the scalar field current 
${\cal J}_{\phi}$ being not conserved as we can see in Eq.~\eqref{back3}. 

\section{Odd-parity and even-parity perturbations}
\label{pertsec}

In this section, we derive the second-order action against BH perturbations 
in the lowest order $U(1)$ GI SVT theories. 
Let us first introduce a metric perturbation $h_{\mu\nu}$ over the background metric 
$\overline{g}_{\mu\nu}$ so that the total metric can be written as $g_{\mu\nu}=\overline{g}_{\mu\nu}+h_{\mu\nu}$. 
The perturbations on top of the spherically symmetric background can be expanded in terms of the spherical harmonics $Y_{lm}(\theta,\varphi)$. 
In the following discussion, we set $m=0$ without loss of generality by virtue of 
the underlying symmetry of the background~\cite{Regge:1957td}. 
Then, the metric perturbation expanded in terms of the $m=0$ modes of the spherical harmonics 
$Y_{l} (\theta)=Y_{l0}(\theta,\varphi)$ can be written as $h_{\mu\nu}=\sum_l h^{(l)}_{\mu\nu}$ 
with the components~\cite{Regge:1957td, Zerilli:1970se, Moncrief:1974gw}, 
\ba
&&
h^{(l)}_{tt}=f(r) H_0 (t,r) Y_{l} (\theta)\,,\qquad 
h^{(l)}_{t r}=h^{(l)}_{r t}=H_1 (t,r) Y_{l} (\theta)\,,\qquad 
h^{(l)}_{t \theta}=h^{(l)}_{\theta t}=h_0 (t,r)
Y_{l,\theta}(\theta)\,,\notag \\
&&
h^{(l)}_{t \varphi}=h^{(l)}_{\varphi t}=-Q (t,r) 
(\sin \theta)Y_{l,\theta}(\theta)\,,\qquad 
h^{(l)}_{rr}=h^{-1}(r) H_2(t,r)Y_{l}(\theta)\,,\qquad 
h^{(l)}_{r \theta}=h^{(l)}_{\theta r}=h_1 (t,r) 
Y_{l,\theta}(\theta)\,,\notag \\
&&
h^{(l)}_{r \varphi}=h^{(l)}_{\varphi r}=-W (t,r) 
(\sin \theta)Y_{l,\theta}(\theta)\,,\qquad 
h^{(l)}_{\theta \theta}=r^2 K(t,r)Y_{l} (\theta)
+r^2 G(t,r)Y_{l,\theta \theta} (\theta)\,,
\notag \\
&&
h^{(l)}_{\theta \varphi}=\frac{1}{2}U(t,r) 
\left[ (\cos \theta)Y_{l,\theta}(\theta)
-(\sin \theta)Y_{l,\theta \theta} (\theta) 
\right]\,,
\notag\\
&&
h^{(l)}_{\varphi \varphi}=r^2 K(t,r)(\sin^2 \theta) 
Y_{l} (\theta)+r^2 G(t,r)
(\sin \theta)(\cos \theta)Y_{l,\theta} (\theta)\,,
\label{metricpert}
\ea
where $Y_{l,\theta} (\theta)=\D Y_{l}(\theta)/\D\theta$, $Y_{l,\theta\theta} (\theta)=\D^2 Y_{l}(\theta)/\D\theta^2$. 
The perturbations can be classified into odd- and even-parity sectors on using the parity transformation $(\theta,\varphi)\to(\pi-\theta,\varphi+\pi)$ 
under which the former changes the sign as $(-1)^{l+1}$ while the latter as $(-1)^l$. 
Among the above metric perturbations, $Q$, $W$, $U$ correspond to the odd-parity modes, 
and $H_0$, $H_1$, $H_2$, $h_0$, $h_1$, $K$, $G$ to the even-parity modes. 
We also introduce the scalar field and the vector field perturbations on top of 
their background contributions $\overline{\phi}$ and $\overline{A}_{\mu}$, 
given in Eqs.~\eqref{phiBG} and \eqref{ABG}, respectively, as follows, 
\be
\phi=\overline{\phi}+\sum_l\delta\phi(t,r)\,Y_l(\theta)\,,\qquad
A_{\mu}=\overline{A}_{\mu}+\sum_l\delta A^{(l)}_{\mu}\,,
\label{scalarpert}
\ee
where 
\be
\delta A_{t}^{(l)}= \delta A_0 (t,r) Y_{l}(\theta)\,,\qquad 
\delta A_{r}^{(l)}= \delta A_1 (t,r) Y_{l}(\theta)\,,\qquad
\delta A_{\theta}^{(l)}= \delta A_2 (t,r) Y_{l,\theta}(\theta)\,,\qquad
\delta A_{\varphi}^{(l)}= -\delta A (t,r) (\sin \theta)Y_{l,\theta}(\theta)\,.
\label{vectorpert}
\ee
The quantity $\delta A$ is classified into the odd-parity sector while $\delta\phi$, $\delta A_0$, $\delta A_1$, $\delta A_2$, correspond to the even-parity modes. 

We consider the infinitesimal gauge transformation $\tilde{x}_{\mu}=x_{\mu}+\sum_l\xi^{(l)}_{\mu}$ with 
\be
\xi^{(l)}_{t}={\cal T}(t,r)Y_{l} (\theta)\,,\qquad
\xi^{(l)}_{r}={\cal R}(t,r)Y_{l} (\theta)\,,\qquad
\xi^{(l)}_{\theta}=\Theta (t,r) Y_{l,\theta} (\theta)\,,\qquad
\xi^{(l)}_{\varphi}=-\Lambda (t,r) (\sin \theta) 
Y_{l,\theta} (\theta)\,,
\label{gaugetrans}
\ee
under which the odd-parity perturbations appearing in Eqs.~\eqref{metricpert}-\eqref{vectorpert} transform as 
\be
\tilde{Q}=Q-\dot{\Lambda}\,,\qquad
\tilde{W}=W-\Lambda'+\frac{2}{r}\Lambda\,,\qquad
\tilde{U}=U-2\Lambda\,,\qquad 
\widetilde{\delta A} =\delta A\,,
\label{oddtrans}
\ee
where a dot represents the derivative with respect to $t$, 
and the even-parity perturbations as 
\ba
&&
\tilde{H}_0=H_0-\frac{2}{f} \dot{\cal T}+\frac{f' h}{f}{\cal R}\,,\qquad 
\tilde{H}_1=H_1-\dot{{\cal R}}-{\cal T}'+\frac{f'}{f}{\cal T}\,,\qquad 
\tilde{H}_2 = H_2-2h{\cal R}'-h' {\cal R}\,,\notag \\
&&
\tilde{h}_0= h_0-{\cal T}-\dot{\Theta}\,,\qquad 
\tilde{h}_1= h_1-{\cal R}-\Theta'+\frac{2}{r} \Theta\,,
\qquad 
\tilde{K} = K-\frac{2}{r}h {\cal R}\,,\qquad
\tilde{G} = G-\frac{2\Theta}{r^2}\,,\qquad 
\widetilde{\delta\phi}=\delta\phi-h\phi'{\cal R}\,,
\notag\\
&&
\widetilde{\delta A}_{0} = 
\delta A_0+\frac{A_0}{f}\dot{{\cal T}} 
-h A_0' {\cal R}\,,\qquad
\widetilde{\delta A}_{1} = 
\delta A_1+\frac{A_0}{f}{\cal T}'
-\frac{f' A_0}{f^2}{\cal T}
\,,\qquad
\widetilde{\delta A}_{2} = 
\delta A_2+\frac{A_0}{f}{\cal T}-h A_1 {\cal R}\,.
\label{eventrans}
\ea
Since the vector field respects the $U(1)$ gauge symmetry, the theory \eqref{act} 
is invariant under the transformation of the vector field perturbation,
\be
\widetilde{\delta A}_{\mu} = \delta A_{\mu}+\partial_{\mu} \sum_{l} \alpha(t,r)Y_{l}(\theta)\,. 
\ee
From Eq.~\eqref{vectorpert}, the perturbations $\delta A_0$, $\delta A_1$, $\delta A_2$ 
transform as 
\be
\widetilde{\delta A}_0=\delta A_0+\dot{\alpha}\,,\qquad
\widetilde{\delta A}_1=\delta A_1+\alpha'\,,\qquad
\widetilde{\delta A}_2=\delta A_2+\alpha\,.
\ee
We can fix the gauge variables $\Lambda$, ${\cal T}$, ${\cal R}$, $\Theta$, and $\alpha$ 
depending on the problem at hand. For instance, we will choose the gauge, 
for the modes with $l=0$ or $l\geq2$, as $\Lambda=U/2$, ${\cal T}=h_0-\dot{\Theta}$, 
${\cal R}=rK/(2h)$, $\Theta=r^2G/2$, and $\alpha=-\delta A_2$. 
This gauge choice eliminates the perturbed quantities as
\be
\tilde{U}=0\,,\qquad
\tilde{h}_0=0\,,\qquad
\tilde{K}=0\,,\qquad 
\tilde{G}=0\,,\qquad 
\widetilde{\delta A}_2=0\,,
\label{gauge}
\ee
and enables one to analyze the dynamics of perturbations more easily. 
We note that some variables in Eq.~\eqref{gauge} automatically vanishes 
in the second-order action without a gauge fixing for $l=1$. 
Hence, one can use the residual gauge DOFs to eliminate other variables 
in addition to those in Eq.~\eqref{gauge} as we will see in Sec.~\ref{dipole}. 
In the following discussion, we apply the gauge choice given in Eq.~\eqref{gauge} 
from the beginning and omit the tilde just for simplicity. 

In the absence of both the magnetic charge $P$ and the parity-violating term $\tF$, 
the odd- and even-parity sectors decouple from each other so that one can analyze 
the dynamics of each sector separately. 
However, the existence of either the magnetic charge or the $\tF$ dependence in $f_2$ 
triggers the mixing of these two sectors~\cite{Kodama:2003kk, Nomura:2020tpc, Pereniguez:2023wxf, Chen:2024hkm}. 
We, then, need to deal with these two sectors together and expand the action~\eqref{act} 
up to second order simultaneously for the odd- and even-parity perturbations 
under the gauge choice \eqref{gauge}. 
After some integration by parts, we obtain the second-order action, 
\be
{\cal S} = \sum_l\int {\rm d}t\, {\rm d}r 
\left({\cal L}_{\rm odd}+{\cal L}_{\rm even}+{\cal L}_{\rm mix}\right)\,,
\label{act2nd}
\ee
with 
\ba
{\cal L}_{\rm odd}&=&
L \Bigg[\alpha_{1} \left(\dot{W} -Q' +\frac{2 Q}{r}\right)^{2}
+2 \left(\alpha_{2} \delta A'+\alpha_{3} {\delta A} \right) 
\left(\dot{W} -Q' +\frac{2 Q}{r}\right)
+\alpha_{4} \dot{\delta A}^{2}+\alpha_{5} \delta A'^{2}\notag\\
&&
+(L -2) \left( \alpha_{6} W^{2}+\alpha_{7} Q^{2}+\alpha_{8} Q {\delta A}\right)
+L \alpha_{9} {\delta A}^{2}+\alpha_{10} W^{2}+\alpha_{11} Q^{2}+\alpha_{12} Q {\delta A}\Bigg]\,,
\label{Lodd}\\
{\cal L}_{\rm even}&=&
H_{0} \Big[a_{1} \delta\phi'' +a_{2} \delta\phi' +a_{3} H_2' +L a_{4} h_1' 
+\left(a_{5}+L a_{6}\right) \delta\phi +\left(a_{7}+L a_{8}\right) H_{2}+L a_{9} h_{1}\Big]
+L b_{1} H_{1}^{2}
\notag\\
&&
+H_{1} \Big(b_{2} \dot{\delta\phi}'+b_{3} \dot{\delta\phi}
+b_{4} \dot{H}_2+L b_{5} \dot{h}_1\Big)+c_{1} \dot{\delta\phi} \dot{H}_2 
+H_{2} \Big[c_{2} \delta\phi' +\left(c_{3}+L c_{4}\right) \delta\phi +L c_{5} h_{1}\Big]
\notag\\
&&
+c_{6} H_{2}^{2}
+L \Big[d_{1} \dot{h}_1^{2}+h_{1} \left(d_{2} \delta\phi'+d_{3} \delta\phi\right)
+d_{4} h_{1}^{2}\Big]+e_{1} \dot{\delta\phi}^{2}+e_{2} \delta\phi'^{2}
+\left(e_{3}+L e_{4}\right) \delta\phi^{2}
\notag\\
&&
+v_{1} \left(\delta A_0' -\dot{\delta A}_1 \right)^{2}
+\left(\delta A_0' -\dot{\delta A}_1 \right) 
\left(v_{2} H_{0}+v_{3} H_{2}+v_{4} \delta\phi'+v_{5} \delta\phi+L v_{6} h_{1}\right)
+\frac{L v_{6} h_{1} \dot{\delta A}_1}{2}
\notag\\
&&
+v_{7} H_{0}^{2}+L \left(v_{8} h_{1} \delta A_0 + v_{9} \delta A_0^{2}
+ v_{10} \delta A_1^{2}+ v_{11} H_{1} \delta A_1 + v_{12} H_{2} \delta A_0 
+ v_{13} \delta\phi \delta A_0\right)\,,
\label{Leven}\\
{\cal L}_{\rm mix}&=&
L \Big[m_{1} H_{0}\delta A +m_{2} H_{2}\delta A +m_{3} 
\left(f \delta A' +A_0' Q \right) h_{1}
+\left(m_{4} \delta A'+m_{5} \delta A +m_{6} Q \right) \delta A_0 
\notag\\
&&
+\left(m_{7} \dot{\delta A} +m_{8} W \right) \delta A_1 
+m_{9} \delta A \delta \phi' 
+\left(m_{10} \delta A'+m_{11} \delta A +m_{12} Q \right) \delta \phi \Big]\,,
\label{Lmix}
\ea
where ${\cal L}_{\rm odd}$ and ${\cal L}_{\rm even}$ correspond to the second-order 
Lagrangians in the odd- and even-parity sectors, respectively, while 
${\cal L}_{\rm mix}$ represents the mixing of these two sectors. 
We have introduced the short-cut notation, 
\be
L=l(l+1)\,, 
\ee
which arises from the integration with respect to the angular coordinates $\theta$ and $\varphi$ 
in the ways that 
\be
\int_0^{2\pi}{\rm d} \varphi \int_0^{\pi} {\rm d} \theta\,
Y_{l, \theta}^2 \sin \theta=L\,,\qquad
\int_0^{2\pi}{\rm d} \varphi \int_0^{\pi} {\rm d} \theta\, 
\left( \frac{Y_{l, \theta}^2}{\sin \theta}+
Y_{l, \theta \theta}^2 \sin \theta \right)=L^2\,. 
\ee
The coefficients $\alpha_1$, $\alpha_2$, ..., $m_{12}$ given in Appendix~\ref{act2nd_coeff} consist solely of background quantities. 
In Eqs.~\eqref{Lodd} and \eqref{Leven}, we have employed the same notation with Ref.~\cite{Zhang:2024cbw} 
in which the stability of hairy BH solutions in the full $U(1)$ gauge-invariant scalar-vector-tensor 
theories without the magnetic charge was studied. Compared to that, there are new terms with the coefficients 
$\alpha_{10}$, $\alpha_{11}$, $\alpha_{12}$ in ${\cal L}_{\rm odd}$, and $m_i$ $(i=1,...,12)$ 
in ${\cal L}_{\rm mix}$ originated from the presence of the magnetic charge and the parity-violating term $\tF$. 
As is obvious, switching off the magnetic charge and the $\tF$-dependence in $f_2$, these new terms identically vanish
and all the coefficients in Eqs.~\eqref{Lodd} and \eqref{Leven} become consistent with those 
in Ref.~\cite{Zhang:2024cbw}. 

\section{Dynamics of perturbation and linear stability conditions}
\label{stabilitysec}

We derive the perturbation equations from the second-order action 
\eqref{act2nd} with Eqs.~\eqref{Lodd}-\eqref{Lmix}, and eliminate the nondynamical 
variables by using them in order to study the linear stability of dyonic BH solutions with a scalar hair. 
In doing so, we deal with the modes $l\geq2$, $l=1$, and $l=0$, separately, 
due to the difference of the number of propagating DOFs. 

\subsection{\texorpdfstring{$l\geq2$}{l>=2}}

There are totally five DOFs in the lowest-order $U(1)$ gauge-invariant SVT theories, i.e., 2 tensor, 2 vector, and 1 scalar DOFs. 
Among them, 1 tensor and 1 vector DOFs appear from the odd mode perturbations while the other 3 from the even mode perturbations. 
In the odd-parity sector Lagrangian~\eqref{Lodd}, the perturbations $W$ and $\delta A$ possess the kinetic terms, namely, $\dot{W}^2$ and $\dot{\delta A}^2$, so that these variables are associated with the dynamical DOFs. 
On the other hand, the variable $Q$, which does not have a kinetic term, should be nondynamical but the existence of the quadratic term $Q'^2$ in Eq.~\eqref{Lodd} prevent one from eliminating this variable by using its perturbation equation in a straightforward way. 
In order to circumvent this problem, we introduce the auxiliary field $\chi$~\cite{DeFelice:2011ka,Kobayashi:2012kh,Kase:2014baa,Heisenberg:2018mgr,Kase:2023kvq,Zhang:2024cbw}
into Eq.~\eqref{Lodd} as follows:
\ba
{\cal L}_{\rm odd}&=&L\Bigg\{
\alpha_1\left[2\chi\left(\dot{W}-Q'
+\frac{2}{r}Q+\frac{\alpha_2\delta A'
+\alpha_3 \delta A}{\alpha_1}\right)-\chi^2\right]
-\frac{(\alpha_2\delta A'+\alpha_3\delta A)^2}
{\alpha_1}
+\alpha_4 \dot{\delta A}^2
+\alpha_5 {\delta A}'^2
\notag\\
&&
+(L-2) \left( \alpha_6 W^2
+\alpha_7 Q^2 
+\alpha_8Q\delta A \right)
+L\alpha_9 \delta A^2
+\alpha_{10}W^2+\alpha_{11}Q^2+\alpha_{12}Q\delta A\Bigg\}\,.
\label{Lodd2}
\ea
The variation of Eq.~\eqref{Lodd2} with respect to the auxiliary field $\chi$ leads 
$\chi=\dot{W}-Q'+(2/r)Q+(\alpha_2\delta A'+\alpha_3\delta A)/\alpha_1$. 
Equation~\eqref{Lodd2} coincides with Eq.~\eqref{Lodd} on the use of this solution that 
guarantees the equivalence of these two Lagrangians. The disappearance of the quadratic term 
$Q'^2$ in the new Lagrangian~\eqref{Lodd2} enables one to eliminate $Q$ by using its 
perturbation equation. We note that, by introducing the auxiliary field $\chi$, 
not only the quadratic term of $Q$ but also the kinetic term of $W$ has been eliminated. 
This implies that the dynamics originally carried by $W$ in Eq.~\eqref{Lodd} should be transferred to 
$\chi$ in Eq.~\eqref{Lodd2} during the process of removing $W$ from the Lagrangian using 
its perturbation equation, as we will see later.

Similarly to the case of $Q$, in the even-parity sector Lagrangian~\eqref{Leven}, the vector field perturbation $\delta A_0$ does not possess a kinetic term $\dot{\delta A_0}^2$, but instead includes a quadratic term $\delta A_0'^2$. 
We introduce the auxiliary field $V$~\cite{Heisenberg:2018mgr,Kase:2023kvq,Zhang:2024cbw} in order to eliminate this quadratic term as 
\ba
{\cal L}_{\rm even}&=&
H_{0} \Big[a_{1} \delta\phi'' +a_{2} \delta\phi' +a_{3} H_2' +L a_{4} h_1' 
+\left(a_{5}+L a_{6}\right) \delta\phi +\left(a_{7}+L a_{8}\right) H_{2}+L a_{9} h_{1}\Big]
+L b_{1} H_{1}^{2}
\notag\\
&&
+H_{1} \Big(b_{2} \dot{\delta\phi}'+b_{3} \dot{\delta\phi}
+b_{4} \dot{H}_2+L b_{5} \dot{h}_1\Big)+c_{1} \dot{\delta\phi} \dot{H}_2 
+H_{2} \Big[c_{2} \delta\phi' +\left(c_{3}+L c_{4}\right) \delta\phi +L c_{5} h_{1}\Big]
+c_{6} H_{2}^{2}
\notag\\
&&
+L \Big[d_{1} \dot{h}_1^{2}+h_{1} \left(d_{2} \delta\phi'+d_{3} \delta\phi\right)
+d_{4} h_{1}^{2}\Big]+e_{1} \dot{\delta\phi}^{2}+e_{2} \delta\phi'^{2}
+\left(e_{3}+L e_{4}\right) \delta\phi^{2}
\notag\\
&&
+v_{1} \left[2V\left(\delta A_0' -\dot{\delta A}_1+\frac{v_{2} H_{0}+v_{3} H_{2}+v_{4} \delta\phi'+v_{5} \delta\phi+L v_{6} h_{1}}{2v_1}\right)-V^2\right]
\notag\\
&&
-\frac{(v_{2} H_{0}+v_{3} H_{2}+v_{4} \delta\phi'+v_{5} \delta\phi+L v_{6} h_{1})^2}{4v_1}
+\frac{L v_{6} h_{1} \dot{\delta A}_1}{2}
\notag\\
&&
+v_{7} H_{0}^{2}+L \left(v_{8} h_{1} \delta A_0 + v_{9} \delta A_0^{2}
+ v_{10} \delta A_1^{2}+ v_{11} H_{1} \delta A_1 + v_{12} H_{2} \delta A_0 
+ v_{13} \delta\phi \delta A_0\right)\,.
\label{Leven2}
\ea
Varying Eq.~\eqref{Leven2} with respect to $V$, we obtain the solution of this auxiliary field as 
$V=\delta A_0' -\dot{\delta A}_1+(v_{2} H_{0}+v_{3} H_{2}+v_{4} \delta\phi'+v_{5} \delta\phi+L v_{6} h_{1})/(2v_1)$ 
which shows the equivalence of the two Lagrangians \eqref{Leven} and \eqref{Leven2}. 
We note that the quadratic term $v_7H_0^2$ in Eq.~\eqref{Leven} is also removed by introducing the auxiliary field $V$. 
Indeed, gathering the coefficients of $H_0^2$ term, we obtain $v_7-v_2^2/(4v_1)$ which identically vanishes on the use of 
relations among $v_1$, $v_2$, and $v_7$, given in Appendix~\ref{act2nd_coeff}. Consequently, the perturbation $H_0$ appears 
only in a linear form in the second-order action showing that $H_0$ is a Lagrange multiplier as in the case of GR 
with $f_2=0$. We will use the relations among the coefficients shown in Appendix~\ref{act2nd_coeff} 
in the following discussion without notification. We also employ the relation 
\be
\left[\frac{2f''}{f}-\frac{f'^2}{f^2}+\frac{(rh'+2h)f'}{rfh}-\frac{2(rh'+2h-2)}{r^2h}\right]a_4
+\frac{4(r^4A_0'^2v_{10}-P^2f^2v_9)}{fr^4}=0\,,
\label{conddf}
\ee
which is derived on the use of the linear combination of background equations, when we remove $f''$ in calculations. 

We now derive the perturbation equations and eliminate the nondynamical variables in order to study the linear stability conditions. 
The variation of the total action \eqref{act2nd} with Eqs.~\eqref{Lmix}, \eqref{Lodd2}, \eqref{Leven2} with respect to 
$W$, $Q$, $\delta A$, $H_0$, $H_1$, $H_2$, $h_1$, $\da_0$, $\da_1$, and $\delta\phi$, we obtain the perturbation equations:
\ba
&&
L \left\{2\left[ (L-2) \alpha_{6} + \alpha_{10}\right] W -2 \alpha_{1} \dot{\chi} +m_{8} \da_1\right\}
=0\,,
\label{eqW}\\
&&
L \left\{2\left[(L-2) \alpha_{7} + \alpha_{11}\right] Q +2 \alpha_{1} \chi' 
+\frac{2 (r\alpha_1' +2 \alpha_{1}) \chi}{r}+\alpha_{12} \da
+m_{3} A_0' h_{1}+m_{6} \da_{0}+m_{12} \delta\phi  \right\}
=0\,,
\label{eqQ}\\
&&
L \Big\{2 \alpha_{3} \chi-2 \alpha_{4} \ddot{\da} -2 \alpha_{5} \da'' -2 \alpha_5' \da' 
+2\left(\alpha_{9} L -\frac{\alpha_{3}^{2}}{\alpha_{1}}\right) \da +\alpha_{12} Q 
+m_{1} H_{0}+m_{2} H_{2}-m_{3} f h_1' 
\notag\\
&&
-(f m_3' +f'm_{3}) h_{1}
-m_{4} \da_0' +(m_{5}-m_4') \da_{0}-m_{7} \dot{\da}_1 +(m_{9}-m_{10}) \delta\phi' 
+(m_{11}-m_{10}') \delta\phi \Big\}
=0\,,
\label{eqdA}\\
&&
\left(a_{2}-\frac{v_{2} v_{4} }{2 v_{1}}\right) \delta\phi'
+a_{3} H_2' +L a_{4} h_1' 
+\left(a_{5}-\frac{v_{2} v_{5} }{2 v_{1}}\right) \delta\phi
-\left(\frac{a_{4} L}{2 h}-a_3'\right) H_{2}+L a_{9} h_{1}
+v_{2}V+L m_{1} \da 
=0\,,
\label{eqH0}\\
&&
2 L b_{1} H_{1}+b_{3} \dot{\delta\phi}+b_{4} \dot{H}_2 +L b_{5} \dot{h}_1
=0\,,
\label{eqH1}\\
&&
- a_{3}H_0'-\frac{L a_{4} H_{0}}{2 h}- b_{4} \dot{H}_1
+\left(c_{2}-\frac{v_{3} v_{4} }{2 v_{1}}\right) \delta\phi'
+\left(c_{3}-\frac{v_{3} v_{5} }{2 v_{1}}\right) \delta\phi
+L c_{5} h_{1}
+\left(2c_{6}-\frac{v_{3}^2}{2 v_{1}}\right) H_2
\notag\\
&&
+v_{3} V+L m_{2} \da
=0\,,
\label{eqH2}\\
&&
L\left[- a_{4}H_0'- (a_4'-a_{9}) H_{0}-  b_{5}\dot{H}_1+  c_{5} H_{2}
-2  d_{1} \ddot{h}_1  + d_{3} \delta\phi+2  d_{4} h_{1} 
+ v_{8} \da_{0}+ m_{3} A_0' Q+ f m_{3} \da'\right]
=0\,,
\label{eqh1}\\
&&
-2 (v_1 V)'
+L\left(
v_{8} h_{1}+2 v_{9} \da_{0}+v_{13} \delta\phi+m_{4} \da'+m_{5} \da  +m_{6} Q
\right)
=0\,,
\label{eqdA0}\\
&&
2 v_{1} \dot{V}+L \left(2 v_{10} \da_1+m_{7} \dot{\da} +m_{8} W\right)
=0\,,
\label{eqdA1}\\
&&
-\left(a_{2}-\frac{v_2v_{4} }{2 v_{1}}\right) H_0'
-b_{3}\dot{H}_1
-\left[\left(c_{2}-\frac{v_3v_{4} }{2 v_{1}}\right) H_2\right]'
+\left(c_{3}-\frac{v_3v_{5} }{2 v_{1}}\right) H_2
+L d_{3} h_{1}
\notag\\
&&
-2 e_{1} \ddot{\delta\phi} 
-\left[\left(2e_{2}-\frac{v_{4}^2 }{2 v_{1}}\right) \delta\phi'\right]'
+\left[2 L e_{4}+2e_3-\frac{v_5^2}{2v_1}+\left(\frac{v_4v_5}{2v_1}\right)'\right] \delta\phi 
\notag\\
&&
-(v_{4} V)'
+v_{5} V 
+L\left[ v_{13} \da_{0}
- (m_{9}-m_{10}) \da' 
+ (m_{11}-m_9') \da 
+ m_{12} Q\right]
=0\,,\label{eqdphi}
\ea
respectively. According to Ref.~\cite{Kobayashi:2014wsa,Kase:2020qvz,Kase:2021mix} 
we introduce the quantity:
\be
\psi=H_2-\frac{L}{r}h_1\,,
\ee
which corresponds to the tensor DOF. We replace $H_2$ and its derivatives by using $\psi$ in what follows. 
We combine Eqs.~\eqref{eqW}, \eqref{eqQ}, \eqref{eqH0}, \eqref{eqH1}, \eqref{eqdA0}, \eqref{eqdA1}, 
and solve them in terms of the nondynamical variables $W$, $Q$, $H_1$, $h_1$, $\da_0$, $\da_1$. 
Substituting those solutions into the second-order action~\eqref{act2nd} with Eqs.~\eqref{Lmix}, \eqref{Lodd2}, \eqref{Leven2}, 
we obtain the reduced action composed only of the dynamical variables $\psi$, $V$, $\delta\phi$, $\chi$, and $\da$, as 
\be
{\cal S}  = \sum_l\int {\rm d}t\, {\rm d}r 
\left(\dot{\vec{\mathcal{X}}}^{t}{\bm K}\dot{\vec{\mathcal{X}}}
+\vec{\mathcal{X}}'^{t}{\bm G}\vec{\mathcal{X}}'
+\vec{\mathcal{X}}'^{t}{\bm S}\vec{\mathcal{X}}
+\vec{\mathcal{X}}^{t}{\bm M} \vec{\mathcal{X}}\right)\,, 
\label{act2nd2}
\ee
where ${\bm K}$, ${\bm G}$, ${\bm M}$ are the $5\times5$ symmetric matrices while ${\bm S}$ is an antisymmetric matrix, 
and the vector $\vec{\mathcal{X}}$ is defined as 
\be
\vec{\mathcal{X}}=\left( 
\begin{array}{c}
\psi\\
\delta \phi\\ 
V\\
\chi\\
\da
\end{array}
\right) \,.
\label{calX}
\ee
As expected, the reduced action \eqref{act2nd2} is composed of the five dynamical variables listed in Eq.~\eqref{calX}, namely, the even-parity perturbations $\psi$, $\delta\phi$, $V$, and the odd-parity perturbations $\chi$, $\delta A$, which is consistent with the fact that the lowest-order $U(1)$ GI SVT theory originally possesses five dynamical DOFs.

\subsubsection{No-ghost conditions}

We demand the kinetic matrix ${\bm K}$ in Eq.~\eqref{act2nd2} to be positive definite for the absence of ghost instabilities.
The matrix components of ${\bm K}$ are given by 
\ba
&&
K_{11}=
\frac{1}{L}\left[{\cal K}_1+\left(\frac{a_4'}{a_4}
+\frac1r+\frac{L}{2rh}\right){\cal K}_2-\frac12{\cal K}_2'\right]\,,\notag\\
&&
K_{12}=
\frac{1}{2L}\left[\frac{fb_3}{ra_4}\left\{{\cal K}_1+\frac12\left(\frac{a_4'}{a_4}
+\frac1r+\frac{L}{2rh}\right){\cal K}_2\right\}
-\frac{1}{ra_4}\left(a_5-\frac{A_0'}{2}v_5\right){\cal K}_2
-\frac12\left(\frac{fb_3}{ra_4}{\cal K}_2\right)'\right]\,,\notag\\
&&
K_{13}=-\frac{A_0'v_1{\cal K}_2}{2Lra_4}\,,\qquad
K_{14}=0\,,\qquad 
K_{15}=\frac{Lm_1}{A_0'v_1}K_{13}\,,
\notag\\
&&
K_{22}=
e_1+\frac{1}{4L}\left[
\frac{fb_3}{r^2a_4^2}
\left\{fb_3{\cal K}_1
-2\left(a_5-\frac{A_0'}{2}v_5\right){\cal K}_2\right\}
-\frac12\left\{\left(\frac{fb_3}{ra_4}\right)^2{\cal K}_2\right\}'
\,\right]\,,\notag\\
&&
K_{23}=\frac{fb_3}{2ra_4}K_{13}\,,\qquad
K_{24}=0\,,\qquad 
K_{25}=\frac{Lfb_3m_1}{2rA_0'a_4v_1}K_{13}\,,
\notag\\
&&
K_{33}=\frac{v_1^2}{L}\left[-\frac{1}{v_{10}}+\frac{2P^2}{(L-2)r^2a_4}\right]\,,\qquad 
K_{34}=\frac{Pv_1}{f(L-2)}\,,\qquad 
K_{35}=-\frac{Lm_4}{2v_1}K_{33}\,,\notag\\
&&
K_{44}=\frac{Lr^2a_4}{2(L-2)f^2}\,,\qquad 
K_{45}=-\frac{LPm_4}{2(L-2)f}\,,\qquad 
K_{55}=L\left\{v_9+\frac{m_4^2}{2}\left[\frac{P^2}{(L-2)r^2a_4}-\frac{1}{2v_{10}}\right]\right\}\,,\qquad 
\label{Kcompts}
\ea
where 
\be
{\cal K}_1=-\frac{2r^2a_4}{f}\,,\qquad 
{\cal K}_2=\frac{4r^3ha_4}{(L-2h)f+rf'h}\,.
\ee
In the absence of the magnetic charge $P$ and the $\tF$-dependence in $f_2$, the cross terms 
between odd- and even-parity sectors, $K_{i4}$ and $K_{i5}$ $(i=1,2,3)$, identically vanish. 
In other words, the mixing of the two sectors goes away and the odd- and even-parity sectors 
decouple from each other. 
If the determinants of all the principal submatrices of ${\bm K}$ are positive, 
the kinetic matrix is positive definite so that the ghost instabilities do not appear. 
This requirement leads the five conditions, 
\ba
&&
k_1\equiv K_{33}=
\frac{v_1^2}{L}\left[-\frac{1}{v_{10}}+\frac{2P^2}{(L-2)r^2a_4}\right]>0\,,
\label{defk1}\\
&&
k_2\equiv K_{11}K_{22}-K_{12}^2=\frac{64 r^2a_{4}^{2} 
\left[(L -2)r^{2}f a_{4} +2 h (P^{2} f^{2} v_{9}-A_0'^{2} r^{4} v_{10})\right]
(\sqrt{fh}{\cal P}_1-a_{4})}{f^{3} L \phi'^{2} (4 L ra_{4} + \sqrt{fh}{\cal P}_2)^{2}}>0\,,
\label{defk2}\\
&&
k_3\equiv K_{44}(K_{11}K_{22}-K_{12}^2)=\frac{Lr^2a_4}{2(L-2)f^2}k_2>0\,,
\label{defk3}\\
&&
k_4\equiv 
(K_{11}K_{22}K_{33}+2K_{12}K_{13}K_{23}-K_{11}K_{23}^2-K_{22}K_{13}^2-K_{33}K_{12}^2)K_{44}-(K_{11}K_{22}-K_{12}^2)K_{34}^2
\notag\\
&&\hspace{.53cm}
=\frac{32r^{4} v_{1}^{2} a_{4}^{3} \left[(L -2) r^{2}a_{4}+2 P^{2}fh v_{9} \right] (\sqrt{fh}{\cal P}_1 -a_{4}) }
{L(L -2) f^{4} \phi'^{2} (-v_{10}) (4L ra_{4} +\sqrt{fh}{\cal P}_2)^{2}}>0\,,
\label{defk4}\\
&&
k_5\equiv {\rm det}\,{\bm K}=
\frac{32 r^{6} a_{4}^{4} v_{1}^{2}(\sqrt{fh}{\cal P}_1-a_{4}) v_{9}}
{f^{4} \phi'^{2}(4 L ra_{4} +\sqrt{fh}{\cal P}_2)^{2} (-v_{10})}>0\,,
\label{defk5}
\ea
where 
\be
{\cal P}_1= -\frac{(r fh' -2f h  -r f' h)}{2 (fh)^{3/2}}a_{4}\,,\qquad 
{\cal P}_2=\frac{4r(rf'h-2fh)}{f^{3/2}\sqrt{h}}a_4\,.
\ee
The inequalities \eqref{defk1}-\eqref{defk5} must hold for any value of $L$ for the absence of ghosts. 
Let us first consider the limit $L\to\infty$. Then, the inequality \eqref{defk1} is satisfied for 
\be
v_{10}<0\,,
\label{NG1}
\ee
under which the inequality \eqref{defk2} is translated to give 
\be
\sqrt{fh}{\cal P}_1-a_{4}>0\,.
\label{NG2}
\ee
Given the above two conditions, the fifth inequality \eqref{defk5} reduces to 
\be
v_9>0\,.
\label{NG3}
\ee
Since the coefficient $a_4=\sqrt{fh}\Mpl^2/2$ is always positive, all the inequalities 
\eqref{defk1}-\eqref{defk5} sufficiently hold for any value of $L$ 
as long as the conditions \eqref{NG1}-\eqref{NG3} are satisfied. 
Hence, \eqref{NG1}-\eqref{NG3} correspond to the no-ghost conditions 
on top of the SSS background of the dyonic BH solutions with the scalar hair 
in the lowest-order $U(1)$ gauge-invariant SVT theories. 

\subsubsection{Radial propagation speeds}

We derive the propagation speeds of the dynamical DOFs, $\psi$, $\delta\phi$, $V$, $\chi$, and $\delta A$, along the radial direction by assuming the solution of the form $\vec{\mathcal{X}}\propto e^{i(\omega t-kr)}$ where $\omega$ and $k$ are the frequency and the wave number, respectively. 
In the limit $\omega\to\infty$ and $k\to\infty$, the reduced action \eqref{act2nd2} 
leads the dispersion relation associated with the radial propagation speed $c_r$ as 
\be
{\rm det} \left( fhc_r^2 {\bm K}+ {\bm G} \right)=0\,. 
\label{eqcr1}
\ee
The matrix components of ${\bm G}$ are given by 
\ba
&&
{G}_{11}=-\frac{f^2{\cal K}_2^2}{16r^4a_4^2}
\left[\frac{1}{v_1}\left(A_0'v_1+\frac{\phi'}{2}v_4\right)^2
-4(rc_5+c_6)
+\frac{r^2}{L}\left(\frac{v_8^2}{v_9}-4d_4\right)
\right]\,,
\notag\\
&&
{G}_{12}=-\frac{2(\sqrt{fh}{\cal P}_1-a_4)}{r\phi'a_4}G_{11}-\frac{f{\cal K}_2}{4r^2a_4}
\left[c_2+\frac{v_4}{2v_1}\left(A_0'v_1+\frac{\phi'}{2}v_4\right)\right]\,,
\notag\\
&&
G_{13}=\frac{v_{10}}{v_9}K_{13}\,,\qquad 
G_{14}=0\,,\qquad 
G_{15}=-\frac{m_1v_{10}{\cal K}_2}{2ra_4v_9}\,,
\notag\\
&&
{G}_{22}=e_2-\frac{v_4^2}{4v_1}-\frac{4(\sqrt{fh}{\cal P}_1-a_4)}{r\phi'a_4}
\left({G}_{12}+\frac{\sqrt{fh}{\cal P}_1-a_4}{r\phi'a_4}{G}_{11}\right)\,,
\notag\\
&&
G_{23}=\frac{v_{10}}{v_9}K_{23}\,,\qquad 
G_{24}=0\,,\qquad 
G_{25}=\frac{fh'-f'h}{fh\phi'}G_{15}\,,
\notag\\
&&
G_{33}=-\frac{v_1^2}{Lv_9}-\frac{2P^2fhv_1^2}{L(L-2)r^2a_4}\,,\qquad 
G_{34}=-\frac{Phv_1}{L-2}\,,\qquad 
G_{35}=-\frac{Lm_4}{2v_1}G_{33}\,,\notag\\
&&
G_{44}=-\frac{Lr^2ha_4}{2(L-2)f}\,,\qquad 
G_{45}=\frac{LPhm_4}{2(L-2)}\,,\qquad 
G_{55}=-fhK_{55}-\frac{L(fhv_9+v_{10})(m_4^2-4v_9v_{10})}{4v_9v_{10}}\,,
\label{Gcompts}
\ea
where we have used the background equations \eqref{back1}-\eqref{back4}. 
Substituting Eqs.~\eqref{Kcompts} and \eqref{Gcompts} into the dispersion relation~\eqref{eqcr1}, 
it reduces to  
\be
-\frac{fhv_1{\cal K}_2^2}{64r^2\phi'^2v_9v_{10}}(c_r^2-1)^2(fhv_9c_r^2+v_{10})^2
\left[8(\sqrt{fh}{\cal P}_1-a_4)v_1c_r^2-\phi'(4c_2v_1+2A_0'v_1v_4+\phi'v_4^2)\right]=0\,,
\label{eqcr2}
\ee
which shows that the propagation speeds of 5 DOFs decouple from each other. 
Among the solutions of Eq.~\eqref{eqcr2}, we find that the two tensor 
propagation speeds squared $c_{r1}^2$ and $c_{r2}^2$ are luminal, i.e., 
\be
c_{r1}^2=c_{r2}^2=1\,,
\label{cr12}
\ee
representing the fact that the action~\eqref{act} does not contain any nonminimal couplings. 
The two vector propagation speeds squared $c_{r3}^2$ and $c_{r4}^2$ are given by 
\be
c_{r3}^2=c_{r4}^2=-\frac{v_{10}}{fhv_9}\,.
\label{cr34}
\ee
The positivity of $c_{r3}^2$ and $c_{r4}^2$ is guaranteed as long as the no-ghost 
conditions \eqref{NG1} and \eqref{NG3} are satisfied. 
If the $Y$-dependence in $f_2$ is absent, a relation $v_{10}=-fhv_9$ holds and 
the vector propagation speed coincides with that of light. In other words, 
the existence of the coupling $Y$ between scalar and vector fields gives rise to 
the modification of radial propagation speeds. 
The remaining solution of Eq.~\eqref{eqcr2} corresponding to the radial propagation 
speed squared for the scalar DOF is written as 
\be
c_{r5}^2=\frac{\phi'(4c_2v_1+2A_0'v_1v_4+\phi'v_4^2)}{8(\sqrt{fh}{\cal P}_1-a_4)v_1}\geq0\,,
\label{noLap1}
\ee
which must be positive for the absence of Laplacian instability along the radial direction. 

\subsubsection{Angular propagation speeds}

In order to derive the propagation speeds along the angular direction, we assume 
the solution of the form $\vec{\mathcal{X}}\propto e^{i(\omega t-l\theta)}$ and 
consider the limit with large $\omega$ and $l$. The dispersion relation 
associated with the angular propagation speed $c_{\Omega}$ is obtained from the reduced action 
\eqref{act2nd2} as 
\be
{\rm det}\left(\frac{fl^2\cosq}{r^2} {\bm K}+ {\bm M}\right)=0\,.
\label{eqco1}
\ee
Although the matrix components of ${\bm M}$ are quite lengthy, 
substituting them along with Eq.~\eqref{Kcompts} into the dispersion relation~\eqref{eqco1}, 
and using the relations among the coefficients provided in Appendix~\ref{act2nd_coeff} and 
Eq.~\eqref{conddf}, yields a rather simple expression for the leading-order contribution 
in the limit $l\gg1$ as 
\ba
&&
-\frac{2L^3a_4^2v_1}{r^6f\phi'^2v_{10}}(\cosq-1)^2
\left[(fv_9\cosq+r^2\alpha_9)(fv_1\cosq+r^2v_{10})-\frac{f(Pfv_9-r^2m_1)^2\cosq}{r^2A_0'^2}\right]
\notag\\
&&
\times\left[(\sqrt{fh}{\cal P}_1-a_4)(\cosq-1)+\frac{(P^2f^2v_9-r^4A_0'^2v_{10})(fhv_9+v_{10})}{2r^2f^2hv_9}
\right]+{\cal O}(L^2)=0\,.
\label{eqco2}
\ea
As in the case of radial propagation, Eq.~\eqref{eqco2} gives 
the angular propagation speeds of tensor DOFs being luminal, i.e., 
\be
c_{\Omega1}^2=c_{\Omega2}^2=1\,.
\label{co12}
\ee
On the other hand, the vector propagation speeds of the odd- and even-parity sectors 
couple with each other due to the existence of the term ${f(Pfv_9-r^2m_1)^2\cosq}/({r^2A_0'^2})$ 
in Eq.~\eqref{eqco2}. This is a consequence of the mixture between the odd- and even-parity 
sectors caused by the presence of the magnetic charge $P$. These solutions, $c_{\Omega3}^2$ and 
$c_{\Omega4}^2$, are given as 
\ba
&&
c_{\Omega3}^2=-\frac12\left[\beta_1+\beta_2+\beta_3-\epsilon\sqrt{(\beta_1-\beta_2)^2+\beta_3(2\beta_1+2\beta_2+\beta_3)}\right]\,,\label{co3}
\\
&&
c_{\Omega4}^2=-\frac12\left[\beta_1+\beta_2+\beta_3+\epsilon\sqrt{(\beta_1-\beta_2)^2+\beta_3(2\beta_1+2\beta_2+\beta_3)}\right]\,,\label{co4}
\ea
where $\epsilon=(\beta_1-\beta_2)/|\beta_1-\beta_2|$, and
\be
\beta_1=\frac{r^2\alpha_9}{fv_9}\,,\qquad 
\beta_2=\frac{r^2v_{10}}{fv_1}\,,\qquad 
\beta_3=-\frac{(Pfv_9-r^2m_1)^2}{r^2fA_0'^2v_1v_9}\,.
\label{defbeta}
\ee
In the absence of magnetic charge $P$, the quantity $\beta_3$ identically vanishes 
and the above solutions reduce to $c_{\Omega3}^2=-\beta_2=-r^2v_{10}/(fv_1)$ and 
$c_{\Omega4}^2=-\beta_1=-r^2\alpha_{9}/(fv_9)$, respectively. Compared to 
Refs.~\cite{Kase:2023kvq,Zhang:2024cbw}, we find that $c_{\Omega3}^2$ is the 
angular propagation speed square of the vector DOF in the even parity-sector 
and $c_{\Omega4}^2$ in the odd-parity sector. 
The positivity of $c_{\Omega3}^2$ and $c_{\Omega4}^2$ requires 
\ba
&&
\beta_1+\beta_2+\beta_3\leq0\,,
\label{noLap2}\\
&&
\beta_1\beta_2\geq0\,,
\label{noLap3}\\
&&
(\beta_1+\beta_2+\beta_3)^2-4\beta_1\beta_2\geq0\,,
\label{noLap4}
\ea
corresponding to conditions for the absence of Laplacian instabilities 
along the angular direction associated with vector DOFs. 
Finally, the positivity condition of the scalar propagating speed square 
obtained as a remaining solution to Eq.~\eqref{eqco2} is 
\be
c_{\Omega5}^2=1-\frac{(P^2f^2v_9-r^4A_0'^2v_{10})(fhv_9+v_{10})}
{2r^2f^2hv_9(\sqrt{fh}{\cal P}_1-a_4)}\geq0\,,
\label{noLap5}
\ee
under which the angular Laplacian instability of scalar DOF disappears. 

\subsection{\texorpdfstring{$l=1$}{l=1}}
\label{dipole}

Let us study the stability of the dipole mode perturbations characterized by $l=1$ ($L=2$). 
In this case, the metric perturbations $K$ and $G$ appear in the second-order action only 
in the form of $K-G$ \cite{Kase:2023kvq} which can be removed by the use of single gauge DOF. 
We employ the choice $\Theta=-(r^2/2)(K-G-2h{\cal R}/r)$ in Eq.~\eqref{eventrans} so as to realize 
$\tilde{K}-\tilde{G}=0$. 
Compared to the gauge choice \eqref{gauge} adopted for the modes $l\geq2$, the gauge  
${\cal R}$ is still unfixed even after the elimination of $\tilde{K}-\tilde{G}$. 
We then fix this gauge DOF as ${\cal R}=\delta\phi/(h\phi')$ which leads $\widetilde{\delta\phi}=0$. 
Moreover, the odd-parity perturbation $U$ does not appear in the second-order action 
without consideration to the gauge choice for $l=1$~\cite{Kase:2023kvq}. 
Then, we can use the gauge $\Lambda$ in Eq.~\eqref{oddtrans} not to eliminate the perturbation $U$ 
but to obtain $\tilde{W}=0$. 
Fixing the remaining gauge DOFs, ${\cal T}$ and $\alpha$, 
in the same way as Eq.~\eqref{gauge}, the following perturbations 
are removed after the gauge transformation, 
\be
\tilde{W}=0\,,\qquad
\tilde{h}_0=0\,,\qquad
\widetilde{\delta\phi}=0\,,\qquad 
\tilde{G}-\tilde{K}=0\,,\qquad 
\widetilde{\delta A}_2=0\,.
\label{gaugedi}
\ee
The second-order action for $l=1$ with this gauge choice can be obtained 
by additionally setting $W=0$ and $\delta\phi=0$ in Eq.~\eqref{act2nd}. 

Since the tensor DOFs do not possess the dipole contribution, 
there are totally three propagating DOFs, namely, two vector and one scalar DOFs, for $l=1$. 
We first eliminate the nondynamical variables by using their perturbation equations 
and derive the reduced action which consists only of the above three propagating DOFs. 
In doing so, we substitute $W=0$ after deriving the perturbation equations. 
Varying the second-order action \eqref{act2nd} represented by Eqs.~\eqref{Lodd}, 
\eqref{Lmix}, \eqref{Leven2} with respect to the nondynamical variables $W$, $Q$, 
$H_0$, $H_1$, $\da_0$, $\da_1$, and substituting $W=0$ and $\delta\phi=0$ after all, 
we obtain the following perturbation equations:
\ba
&&
\frac{v_{10} P \da_1}{r^{2}}+\dot{\cal E}=0\,,
\label{eqWdi}\\
&&
\frac{\alpha_{12} \da}{2}+\frac{P v_{9} \da_0}{r^{2}}+\frac{v_{10} P A_0' h_{1}}{f r^{2}}+\frac{P^{2} v_{9} Q}{r^{4}}-\frac{1}{r^{2}}(r^2{\cal E})'=0\,,
\label{eqQdi}\\
&&
r a_{4} \psi' -2 m_{1} \da +\frac{\left(f r h' +r h f' +2 f h +2 f \right) a_{4} \psi}{2 f h}-A_0' v_{1} V -\frac{a_{4} \left(-r h f' +2 f h -2 f \right) h_{1}}{r f h}=0\,,
\label{eqH0di}\\
&&
r a_{4} \dot{\psi} +a_{4} \dot{h}_1 +a_{4} H_{1}=0\,,
\label{eqH1di}\\
&&
\frac{2 A_0' v_{10} h_{1}}{f}+2 v_{9} \da_0+\frac{2 \left(f P v_{9}-m_{1} r^{2}\right) \da'}{r^{2} A_0'}+m_{5} \da +\frac{2 P v_{9} Q}{r^{2}}-(v_{1}V)'=0\,,
\label{eqdA0di}\\
&&
-\frac{2 \left(f P v_{9}-m_{1} r^{2}\right) \dot{\da}}{r^{2} A_0'}+2 v_{10} \da_1+v_{1} \dot{V}=0\,,
\label{eqdA1di}
\ea
respectively, where we have introduced the combination 
\be
{\cal E}=\alpha_1\left(Q'-\frac{2Q}{r}-\frac{\alpha_3}{\alpha_1}\da\right)\,.
\label{defEp}
\ee
In the absence of the magnetic charge, Eqs.~\eqref{eqWdi} and \eqref{eqQdi} simply 
reduce to $\dot{\cal E}=0$ and $r^{-2}(r^2{\cal E})'=0$ which are obviously integrable. 
However, the mixing between the odd- and even-parity sectors caused by the existence of 
the magnetic charge $P$ gives rise to the several new terms in Eqs.~\eqref{eqWdi} and \eqref{eqQdi} 
which make the structure of the odd-parity perturbations different from the case without $P$, and 
these perturbation equations cannot be simply integrated at this step. 
We then combine Eqs.~\eqref{eqH0di}-\eqref{eqdA1di} and solve them in terms of 
$h_1$, $H_1$, $\da_0$, $\da_1$. Substituting these solutions into 
Eqs.~\eqref{eqWdi} and \eqref{eqQdi}, we obtain 
\be
\frac{\D}{\D t}\left[\frac{P(Pfv_9-r^2m_1)\da}{r^4A_0'}-\frac{Pv_1V}{2r^2}+{\cal E}\right]=0\,,
\qquad
\frac{\D}{\D r}\left[\frac{P(Pfv_9-r^2m_1)\da}{r^2A_0'}-\frac{Pv_1V}{2}+r^2{\cal E}\right]=0\,,
\ee
respectively. These equations now can be straightforwardly integrated to give the solution 
\be
{\cal E}=\alpha_1\left(Q'-\frac{2Q}{r}-\frac{\alpha_3}{\alpha_1}\da\right)=
\frac{Pv_1V}{2r^2}-\frac{P(Pfv_9-r^2m_1)\da}{r^4A_0'}+\frac{\cal C}{r^2}\,,
\label{solEp}
\ee
where ${\cal C}$ is a constant. We also substitute the solutions of 
$h_1$, $H_1$, $\da_0$, $\da_1$ into the second-order action represented 
by Eqs.~\eqref{Lodd}, \eqref{Lmix}, \eqref{Leven2}, and find that 
the perturbation $Q$ and its derivatives appear only in the form $Q'-2Q/r$ 
in the resultant action. Removing $Q'-2Q/r$ on the use of Eq.~\eqref{solEp}, 
we eventually obtain the reduced action as in the same form as Eq.~\eqref{act2nd2}
with the dynamical DOFs, 
\be
\vec{\mathcal{X}}=\left( 
\begin{array}{c}
\psi\\
V\\
\da
\end{array}
\right) \,.
\label{calXdi}
\ee
We note that the quantity $\psi$ does not correspond to the tensor DOF 
since it is absent in the dipolar perturbations. Instead, $\psi$ represents 
the scalar DOF embedded via the gauge choice \eqref{gaugedi}. 

The no-ghost conditions for the dipole mode are absent under the conditions 
\be
K_{22}>0\,,\qquad 
K_{22}K_{33}-K_{23}^2>0\,,\qquad 
{\rm det} \bm{K}>0\,.
\ee
The first inequality reduces to \eqref{NG1} under which the second leads \eqref{NG3}. 
Under these conditions, the third inequality reduces to \eqref{NG2}. The radial 
propagation speeds derived from the dispersion relation~\eqref{eqcr1} for $l=1$ 
coincide with those in Eqs.~\eqref{cr34} and \eqref{noLap1}. Hence, the two vector and 
one scalar DOFs for the mode $l=1$ obey the same stability conditions for the modes $l\geq2$. 

\subsection{\texorpdfstring{$l=0$}{l=0}}
\label{monopole}

We consider the monopole perturbation characterized by $l=0$ ($L=0$) which possesses 
only one propagating DOF, the scalar field perturbation. 
For $l=0$, the second-order Lagrangian in the odd-parity sector given in Eq.~\eqref{Lodd} 
and the mixing terms in Eq.~\eqref{Lmix} identically vanish. The perturbations $h_0$, $h_1$, 
and $G$ in the even-parity sector Lagrangian given in Eq.~\eqref{Leven} also vanish identically. 
This shows that, regarding the gauge transformation, the perturbations $\tilde{U}$, $\tilde{h}_0$, 
and $\tilde{G}$, are removed from the second-order action without using the gauge DOFs as in 
Eq.~\eqref{gauge}. Although one can fix the corresponding gauge DOFs in order to eliminate 
other variables, such a gauge choice results in an incomplete gauge fixing. 
Thus, we adopt the same gauge choice as Eq.~\eqref{gauge} for the mode with $l=0$. 

Varying the action \eqref{act2nd} with Eq.~\eqref{Leven2} with respect to $H_0$, $H_1$, 
$\da_0$, $\da_1$, we obtain 
\ba
&&
\Phi'+A_0'v_1V=0\,,
\label{eqH0mo}\\
&&
\dot{\Phi}=0\,,
\label{eqH1mo}\\
&&
(v_1V)'=0\,,
\label{eqdA0mo}\\
&&
\dot{V}=0\,,
\label{eqdA1mo}
\ea
respectively. Here, we have introduced the combination 
\be
\Phi=-\frac{f}{2}(b_3\delta\phi+b_4H_2)\,.
\label{defPhi}
\ee
Equations~\eqref{eqdA0mo} and \eqref{eqdA1mo} are integrated to give 
\be
V=\frac{{\cal C}_1}{v_1}\,,
\label{solV}
\ee
where ${\cal C}_1$ is a constant. Substituting this solution into 
Eq.~\eqref{eqH0mo} and integrating it with Eq.~\eqref{eqH1mo}, 
we obtain  
\be
\Phi={\cal C}_2-{\cal C}_1A_0\,,
\ee
where ${\cal C}_2$ is a constant. Combining the definition of $\Phi$ given in 
Eq.~\eqref{defPhi} with this solution, the perturbation $H_2$ can be written as 
\be
H_2=-\frac{b_3}{b_4}\delta\phi-\frac{2}{fb_4}({\cal C}_2-{\cal C}_1A_0)\,.
\label{solH2}
\ee
Besides the perturbation $H_0$ for any $l$, the perturbations $H_1$, $\da_0$, 
and $\da_1$ for the monopole mode $l=0$ do not possess the quadratic terms 
in the second-order Lagrangian \eqref{Leven2}. This shows that these quantities 
are Lagrange multipliers and they are completely removed from the second-order action 
by substituting the solutions \eqref{solV} and \eqref{solH2} of their perturbation 
equations~\eqref{eqH0mo}-\eqref{eqdA1mo}. This process also replaces the perturbation $H_2$ 
with the dynamical variable $\delta\phi$ through Eq.~\eqref{solH2}. Omitting the integration 
constants ${\cal C}_1$ and ${\cal C}_2$ which are irrelevant for the stability conditions, 
we obtain  the reduced action composed only of the dynamical DOF, $\delta\phi$, as
\ba
&&{\cal S}_{l=0} = \int {\rm d}t\, {\rm d}r\Bigg[
e_1\dot{\delta\phi}^2-\frac{v_4^2-4v_1e_2}{4v_1}\delta\phi'^2
-\frac{1}{2v_1}\left\{v_4v_5+\frac{b_3(2c_2v_1-v_3v_4)}{b_4}\right\}\delta\phi\delta\phi'
\notag\\
&&\hspace{2.5cm}
+\frac{1}{b_4^2}\left\{b_3^2c_6-b_3b_4c_3+b_4^2e_3-\frac{(b_3v_3-b_4v_5)^2}{4v_1}\right\}\delta\phi^2
\Bigg]\,.
\label{act2ndmo}
\ea
The absence of the ghost requires the coefficient of the kinetic term $\dot{\delta\phi}^2$ being positive, i.e.,
\be
e_1=\frac{2(\sqrt{fh}{\cal P}_1-a_4)}{fh\phi'^2}>0\,,
\ee
which is consistent with the no-ghost condition \eqref{NG2} derived for the mode $l\geq2$. 
The radial propagation speed square of $\delta\phi$ is given by 
\be
c_r^2=\frac{1}{fh}\frac{v_4^2-4v_1e_2}{4v_1e_1}
=\frac{\phi'(4c_2v_1+2A_0'v_1v_4+\phi'v_4^2)}{8(\sqrt{fh}{\cal P}_1-a_4)v_1}\,
\ee
which coincides with that for $l\geq2$ given in Eq.~\eqref{noLap1}. These results ensure 
the monopole perturbation being stable under the stability conditions for $l\geq2$. 

In conclusion, the individual conditions derived in this section are 
three no-ghost conditions \eqref{NG1}, \eqref{NG2}, \eqref{NG3}, 
one condition for the absence of Laplacian instability along the radial direction \eqref{noLap1}, 
four conditions for the absence of Laplacian instabilities along the angular direction 
\eqref{noLap2}, \eqref{noLap3}, \eqref{noLap4}, \eqref{noLap5}. Namely,
\ba
{\rm No\ ghosts}&:&\qquad 
v_9>0\,,\qquad 
v_{10}<0\,,\qquad 
\sqrt{fh}{\cal P}_1-a_4>0\,,
\notag\\
c_r^2>0&:&\qquad 
\phi'(4c_2v_1+2A_0'v_1v_4+\phi'v_4^2)v_1>0\,,
\notag\\
\cosq>0&:&\qquad 
\beta_1+\beta_2+\beta_3\leq0\,,\qquad
\beta_1\beta_2\geq0\,,\qquad 
(\beta_1+\beta_2+\beta_3)^2-4\beta_1\beta_2\geq0\,,\notag\\
&&\hspace{.75cm}
1-\frac{(P^2f^2v_9-r^4A_0'^2v_{10})(fhv_9+v_{10})}
{2r^2f^2hv_9(\sqrt{fh}{\cal P}_1-a_4)}\geq0\,.
\label{stability}
\ea
%

\section{Application to the concrete models}
\label{modelsec}

In this section, we apply our general stability conditions \eqref{stability} derived in 
Sec.~\ref{stabilitysec} to the concrete models which possess hairy BH solutions. We focus 
on extended Einstein-Maxwell-scalar theories~\cite{Taniguchi:2024ear} described by, 
\be
f_2=F+X+g_1(\phi,X)\,F+g_2(\phi,X)\,\tilde{F}+\bar{g}_3(\phi,X)\,Y\,.
\label{eEMS}
\ee
The derivative interaction $Y$ in the absence of the magnetic charge $P$ 
reduces to $\overline{Y}=4\overline{X}\overline{F}$ on the background 
as we can see in Eq.~(\ref{XBG}). 
Although this relation does not hold for $P\neq0$, the quantity $\overline{Y}$ 
is still proportional to $\overline{X}$ as $\overline{Y}=(2hA_0'^2/f)\overline{X}$. 
Assuming that $\phi$ and $A_0$ are regular on the horizon characterized by $f=0=h$, 
the above relation implies that the interaction term $\bar{g}_3Y$ can be regular 
even for the coupling $\bar{g}_3\propto X^{-1}$. 
The authors in Ref.~\cite{Taniguchi:2024ear} introduced the normalization of 
the coupling $\bar{g}_3$ in order to incorporate the above discussion explicitly as
\be
\bar{g}_3(\phi,X)=\frac{g_3(\phi,X)}{4X}
\ee
without loss of generality. 

The extended Einstein-Maxwell-scalar theories accommodate the several concrete models 
known in the literature to have hairy BH solutions through the scalar type interaction $g_1(\phi)F$~\cite{Gibbons:1987ps,Gibbons:1990um,Garfinkle:1990qj,Sheykhi:2014gia,Herdeiro:2018wub,Astefanesei:2019pfq,Fernandes:2020gay}, 
and the axionic-type interaction $g_2(\phi)\tF$~\cite{Lee:1991jw,Lee:1991qs,Boskovic:2018lkj,Fernandes:2019kmh,Herdeiro:2019tmb}. 
In Ref.~\cite{Taniguchi:2024ear}, the new type of hairy BH solutions arising from 
the derivative interaction $\bar{g}_3(\phi,X)Y$ are discovered, and the effect of $X$-dependence 
in the functions $g_1$ and $g_2$ is also investigated. 

As we have studied in Sec.~\ref{stabilitysec}, the radial and angular propagation speeds 
of tensor DOFs given in Eqs.~\eqref{cr12} and \eqref{co12}, respectively, are luminal 
irrespective of the functional form of $f_2$, i.e., 
\be
c_{r1}^2=c_{r2}^2=c_{\Omega1}^2=c_{\Omega2}^2=1\,. 
\ee
Moreover, in the extended Einstein-Maxwell-scalar theories described by Eq.~\eqref{eEMS}, 
we have $\beta_1=-1$, $\beta_2=-1$, and $\beta_3=0$ in Eq.~\eqref{defbeta} which result in 
\be
c_{\Omega3}^2=c_{\Omega4}^2=1\,,
\ee%
from Eqs.~\eqref{co3} and \eqref{co4}. We note that the positivity of radial propagation 
speed squares associated with vector DOFs, $c_{r3}^2$ and $c_{r4}^2$ , are ensured under 
the no-ghost conditions as discussed below Eq.~\eqref{cr34}. Accordingly, only five 
non-trivial conditions are left for the linear stability of hairy BHs in extended the 
Einstein-Maxwell-scalar theories. They are the three no-ghost conditions \eqref{NG1}, 
\eqref{NG2}, \eqref{NG3}, and the two conditions for the absence of Laplacian instabilities 
associated with the scalar DOF along radial and angular directions given by 
\eqref{noLap1} and \eqref{noLap5}, respectively. These conditions reduce to 
\ba
&&
s_1\equiv1+g_1+g_3>0\,,\label{con1}
\\
&&
s_2\equiv1+g_{1,X}\overline{F}+g_{2,X}\overline{\tF}+\frac{g_{3,X}\overline{Y}}{4\overline{X}}>0\,,\label{con2}
\\
&&
s_3\equiv1+g_1>0\,,\label{con3}
\\
&&
c_{r5}^2=1+\frac{h\phi'^2}{s_1s_2}\Bigg[\left\{\sqrt{\frac{h}{f}}A_0'(g_{1,X}+g_{3,X})-\frac{P}{r^2} g_{2,X}\right\}^2-s_1\left(g_{1,XX}\overline{F}+g_{2,XX}\overline{\tF}+\frac{g_{3,XX}\overline{Y}}{4\overline{X}}\right)\Bigg]
\geq0\,,
\label{con4}\\
&&
c_{\Omega5}^2=1+\frac{g_3(r^4hA_0'^2s_1+P^2f s_3)}{r^4fh\phi'^2s_2s_3}\geq0\,,
\label{con5}
\ea
respectively. 
In the following, we focus on the three cases characterized by 
(i) $g_1\neq0$, $g_2=0$, $g_3=0$, 
(ii) $g_1=0$, $g_2\neq0$, $g_3=0$, 
(iii) $g_1=0$, $g_2=0$, $g_3\neq0$, 
for the sake of concreteness. 

\subsection*{\texorpdfstring{(i) $g_1\neq0$, $g_2=0$, $g_3=0$}{case (i)}}

The stability conditions \eqref{con1}-\eqref{con5} in the case (i) reduce to 
\ba
&&
s_1=s_3=1+g_1>0\,,\label{con1a}\\
&&
s_2=1+g_{1,X}\overline{F}>0\,,\label{con2a}\\
&&
c_{r5}^2=1+\frac{h\phi'^2\left(hA_0'^2g_{1,X}^2/f-s_1g_{1,XX}\overline{F}\right)}{s_1s_2}\geq0\,,
\label{con3a}
\ea
and $c_{\Omega5}^2=1$. Thus, the five stability conditions converge into three. 
For the coupling $g_1(\phi)F$~\cite{Gibbons:1987ps,Gibbons:1990um,Garfinkle:1990qj,Sheykhi:2014gia,Herdeiro:2018wub,Astefanesei:2019pfq,Fernandes:2020gay}, the second and the third conditions \eqref{con2a}-\eqref{con3a} 
are automatically satisfied, and the stability of the model only requires the first condition \eqref{con1a}. 
Let us consider the Einstein-Maxwell-dilaton theory, for instance, in which an exact hairy BH solution 
discovered by Gibbons and Maeda (GM) \cite{Gibbons:1987ps,Gibbons:1990um}, and Garfinkle, Horowitz, 
and Strominger (GHS) \cite{Garfinkle:1990qj} exists. This theory is described by the action 
in the Einstein frame as
\be
S = \int {\rm d}^4 x \sqrt{-g} 
\left( R -2\nabla_{\mu}\Phi\nabla^{\mu}\Phi +4e^{-2\Phi} F\right)\,.
\ee
In the extended Einstein-Maxwell-scalar theories, the above model is reproduced by 
redefining $\phi=2\Phi$ and setting the functional form $1+g_1=4e^{-\phi} (=4e^{-2\Phi})$ with the unit $\Mpl^2/2=1$,
which satisfies the only stability condition \eqref{con1a} automatically. 
We note that the stability of GMGHS solution dressed in an electric charge alone was
confirmed in Ref.~\cite{Kase:2023kvq}. In this paper, we discover that 
not only electrically but also magnetically charged GMGHS solution is stable 
against linear perturbations. 

\subsection*{\texorpdfstring{(ii) $g_1=0$, $g_2\neq0$, $g_3=0$}{case (ii)}}

In this case, the stability conditions \eqref{con1}, \eqref{con3}, and \eqref{con5} 
are automatically satisfied since they reduce to $s_1=1$, $s_3=1$, and $c_{\Omega5}^2=1$, 
respectively. The remaining conditions \eqref{con2} and \eqref{con4} are given as 
\be
s_2=1+g_{2,X}\overline{\tF}>0\,,\qquad 
c_{r5}^2=1+\frac{h\phi'^2\left(P^2g_{2,X}^2/r^4-g_{2,XX}\overline{\tF}\right)}{s_2}\geq0\,,
\label{con1b}
\ee
respectively. The coupling $g_2(\phi)\tF$~\cite{Lee:1991jw,Lee:1991qs,Boskovic:2018lkj,Fernandes:2019kmh,Herdeiro:2019tmb} 
trivially satisfies the above two conditions as $s_2=1$ and $c_{r5}^2=1$. Thus, we conclude 
that the hairy BH solutions arising from the axionic-type coupling $g_2(\phi)\tF$ are 
inevitably stable regardless of the functional form of $g_2(\phi)$. 

\subsection*{\texorpdfstring{(iii) $g_1=0$, $g_2=0$, $g_3\neq0$}{case (iii)}}

The stability conditions given in Eqs.~\eqref{con1}-\eqref{con5} are translated to 
the following four nontrivial conditions, 
\ba
&&
s_1=1+g_3>0\,,\label{con1c}\\
&&
s_2=1+\frac{hA_0'^2}{2f}g_{3,X}>0\,,\label{con2c}\\
&&
c_{r5}^2=1+\frac{h\phi'^2}{s_1s_2}\left(\frac{hA_0'^2}{f}g_{3,X}^2
-\frac{s_1\overline{Y}}{4\overline{X}}g_{3,XX}\right)\geq0\,,\label{con3c}\\
&&
c_{\Omega5}^2=1+\frac{g_3(r^4hA_0'^2s_1+P^2f)}{r^4fh\phi'^2s_2}\geq0\,,\label{con4c}
\ea
with the trivial condition $s_3=1>0$ in the case (iii). In the absence of the $X$-dependence 
in $g_3$, the second and the third conditions reduce to $s_2=1>0$ and $c_{r5}^2=1\geq0$ 
while the first and the fourth conditions remain nontrivial. In Ref.~\cite{Taniguchi:2024ear}, 
the authors showed that the concrete coupling $g_3=c_3\phi/\Mpl$ gives rise to a new hairy solution 
where $c_3$ is a constant. The remaining two stability conditions in this concrete model are 
sufficiently satisfied for the positive parameter $c_3 \phi >0$. 

\section{Conclusions}
\label{concludesec}

We studied linear perturbations of dyonic BH solutions with a scalar hair in the lowest-order $U(1)$ GI SVT theories described by the action \eqref{act}. 
The perturbations on top of the SSS background given in Eq.~\eqref{line} can be classified into odd-parity and even-parity perturbations, depending on their transformation properties under a parity transformation. 
While the odd-parity and even-parity sectors completely decouple from each other in the second-order action for perturbations in the case of BHs with purely electric charges, a mixing of these sectors occurs in the case of dyonic BH solutions dressed in not only the electric charge but also a magnetic charge represented as a quantity $P$ in Eq.~\eqref{ABG}. 
We have expanded the action in terms of both odd-parity and even-parity perturbations simultaneously, and derived the perturbation equations as well as the general conditions for the stability of dyonic BH solutions. 
We represent the first attempt to analyze the stability of hairy black hole solutions containing a scalar field coupled to a vector field carrying both electric and magnetic charges, in terms of the action level.
Our new framework enables us to discuss the stability of dyonic hairy BH solutions arising from various theories within the subclass of the lowest-order $U(1)$ GI SVT theories.
We also applied these general conditions to the concrete models which give rise to the dyonic BH solutions with the scalar hair and examined their stabilities. 

In Sec.~\ref{pertsec}, we introduced the perturbations on top of the SSS background as in Eqs.~\eqref{metricpert}-\eqref{vectorpert}. 
Among 15 perturbation variables, 4 of them are classified into the odd-parity sector while the remaining 11 are classified into the even-parity sector. 
The infinitesimal gauge transformation \eqref{gaugetrans} associated with a general covariance of the theory enables one to eliminate three even-parity perturbations as well as one odd-parity perturbation. 
Moreover, the existence of $U(1)$ gauge invariance in the theory enables us to eliminate one more even-parity variable arising from vector field perturbation. 
We fixed the gauge DOFs as in Eq.~\eqref{gauge} to eliminate 5 perturbed variables, and expanded the action \eqref{act} up to second order in terms of the remaining 10 perturbations. 
The resultant second-order action given in Eq.~\eqref{act2nd} is expressed by ${\cal L}_{\rm odd}$, ${\cal L}_{\rm even}$, and ${\cal L}_{\rm mix}$, which represent the Lagrangian densities for the odd-parity sector, the even-parity sector, and the mixing of these two sectors, respectively.

In Sec.~\ref{stabilitysec}, we derived perturbation equations of the 10 variables as in Eqs.~\eqref{eqW}-\eqref{eqdphi}. 
For the multipole modes $l\geq2$, we removed 5 nondynamical variables from the second-order action by using the perturbation equations, and obtained the reduced action \eqref{act2nd2} composed of 5 dynamical DOFs in Eq.~\eqref{calX} which coincide with the number of propagation DOFs in the theory \eqref{act}, i.e., two tensors, two vectors, and one scalar.  
Due to the mixing of odd- and even-parity sectors originated from the magnetic charge, the reduced action contains the nonzero cross terms between these two sectors. 
We, then, derived the no-ghost conditions \eqref{defk1}-\eqref{defk5} as well as the propagation speeds along the radial direction \eqref{cr12}-\eqref{noLap1} and those along the angular direction \eqref{co12}, \eqref{co3}, \eqref{co4}, \eqref{noLap5} which must be positive for the absence of Laplacian instabilities. 
After excluding trivial and redundant conditions from the general ones, we are left with eight independent constraints, as shown in Eq.~\eqref{stability}. 
We also examined the monopole ($l=0$) and the dipole ($l=1$) modes in Sec.~\ref{dipole} and \ref{monopole}, respectively, and confirmed that their stability is guaranteed, as long as the stability conditions for the multipole ($l\geq2$) modes are satisfied. 

In Sec.~\ref{modelsec}, we applied the general stability conditions derived in Sec.~\ref{stabilitysec} to the concrete models which possess the dyonic BH solutions with a scalar hair. 
We focused on extended Einstein-Maxwell-scalar theories described by Eq.~\eqref{eEMS} which accommodate the several concrete models known in the literature. 
In these theories, some of the stability conditions are trivially satisfied, and only five non-trivial conditions \eqref{con1}-\eqref{con5} are left. 
For the scalar-vector coupling of the form $g_1(\phi,X)F$, these non-trivial conditions reduce to Eqs.~\eqref{con1a}-\eqref{con1c}. 
If the coupling function $g_1$ depends only on $\phi$, the model is stable as long as $1+g_1>0$. 
The result that this inequality is automatically satisfied even in the case of the GMGHS solution is one of the key achievements of this paper.
The coupling of the form $g_2(\phi,X)\tF$ leads to two independent conditions~\eqref{con1b}. 
Interestingly, for the axionic interaction $g_2(\phi)\tF$, all the stability conditions become trivial.
Hence, all the axionic BHs originated from such coupling are stable against linear perturbations.
We also investigated the derivative interaction of the form $\bar{g}_3(\phi,X)Y$. 
In this case, the four conditions \eqref{con1c}-\eqref{con4c} are required for the stability of the model. 
Focusing on the specific coupling $g_3 = c_3\phi/\Mpl$, which was found in Ref.~\cite{Taniguchi:2024ear} to generate new hairy solutions, the positivity of $c_3 \phi$ serves as a sufficient condition for stability.
These analytical results support the exploration of BH solutions which are both hairy and stable.

In this paper, we constructed the general framework to study the dynamics of perturbations on top of the SSS background for dyonic BHs with a scalar hair in the lowest-order $U(1)$ GI SVT theories
\footnote{The analysis involves theory-dependent steps that prevent full automation, but the procedure and results presented here ensure reproducibility of the stability conditions.}.
We applied our general stability conditions to the extended Einstein-Maxwell-scalar theory as a concrete example, which contains only linear terms of $F$, $\tF$, and $Y$. 
If we consider the nonlinear Lagrangian of the form ${\cal L}(F)$, the non-singular BH solutions exist \cite{Ayon-Beato:1998hmi,Ayon-Beato:1999kuh,Ayon-Beato:2000mjt}, but it is shown that such solutions are prone to Laplacian instabilities even if we consider the extended Lagrangian ${\cal L}(F,\phi,X)$ in Ref.~\cite{DeFelice:2024ops}. 
There would be a possibility to construct a model possessing the stable and non-singular BH solutions within our more general setup $f_2(\phi,X,F,\tilde{F},Y)$. 
The perturbation equations given in Eqs.~\eqref{eqW}-\eqref{eqdphi} provide a basis for in-depth research related BH physics, e.g., the computation of quasinormal modes \cite{Vishveshwara:1970zz, Press:1971wr, Kokkotas:1999bd, Nollert:1999ji}, the evaluation of Hawking radiation \cite{Hawking:1974rv, York:1983zb}, the exploration of BH superradiance \cite{Klein:1929zz, Dicke:1954zz, ZelDovich:1971, ZelDovich:1972, Bekenstein:1973mi, Bekenstein:1998nt, Brito:2015oca}, and the comparison between our linear stability conditions and BH thermodynamic stability conditions \cite{Hawking:1976de, Chamblin:1999hg}.
It is also of interest to extend our analysis to the full $U(1)$ GI SVT theories~\cite{Heisenberg:2018acv} after exploring background solutions in the presence of a magnetic charge. 
These issues are left for future work. 

\section*{Acknowledgments}
R K. is supported by the Grant-in-Aid for Scientific Research (C) of the JSPS No.~23K03421. 

\section*{Data Availability}
The data are not publicly available. The data are available from the authors upon reasonable request.

\appendix

\section{Coefficients in the second-order action}
\label{act2nd_coeff}

The coefficients in the second-order action \eqref{act2nd} 
with Eqs.~\eqref{Lodd}-\eqref{Lmix} are given as follows: 
\ba
&&
\alpha_1=\frac{a_4}{2f}\,,\qquad
\alpha_2=0\,,\qquad 
\alpha_3=\frac{A_0'}{f}v_{10}\,,\qquad 
\alpha_4=v_9\,,\qquad 
\alpha_5=v_{10}\,,\qquad
\alpha_6=-\frac{a_4}{2r^2}\,,
\notag\\
&&
\alpha_7=\frac{a_4}{2r^2fh}\,,\qquad
\alpha_8=0\,,\qquad
\alpha_9=
-\frac{\sqrt{f} \left(r^{4} f_{2,F}-P^{2} f_{2,{FF}}\right)}{2 r^{6} \sqrt{h}}
+\frac{P {A_0'} f_{2,F {\tF}}}{r^{4}}
+\frac{\sqrt{h} {A_0'}^{2} f_{2,{\tF\tF}} }{2 r^{2} \sqrt{f}} 
\,,\notag\\
&&
\alpha_{10}=\frac{P^2}{r^4}v_{10}\,,\qquad
\alpha_{11}=\frac{P^2}{r^4}v_9\,,\qquad 
\alpha_{12}=\frac{P}{r^2}(f_{2,\tF})'\,,\qquad 
a_1=0\,,\qquad
a_2=\frac{r \left(f'h - fh' \right) a_{4}}{fh\phi'}+\frac{A_0' v_{4}}{2}\,,
\notag\\
&&
a_3=-ra_4\,,\qquad 
a_4=\frac{\sqrt{fh}\Mpl^{2}}{2}\,,\qquad
a_5=a_2'-\frac{A_0''}{2}v_4-\frac12(v_4'-v_5)A_0'\,,\qquad
a_6=0\,,
\notag\\
&&
a_7=-\left(1+\frac{r f'}{2 f}+\frac{r h'}{2 h}\right) a_{4}
-\frac14{A_0' \left(2 A_0' v_{1}+\phi' v_{4}\right)}\,,\qquad
a_8=-\frac{a_4}{2h}\,,\qquad 
a_9=\frac{\left(r h' +2 h \right)}{2 h r}a_{4}\,,
\notag\\
&&
b_1=\frac{a_4}{2f}\,,\qquad
b_2=0\,,\qquad 
b_3=-\frac{2r(f'h-fh')}{f^2h\phi'}a_4\,,\qquad 
b_4=\frac{2r}{f}a_4\,,\qquad
b_5=-\frac{a_4}{f}\,,\qquad 
c_1=0\,,
\notag\\
&&
c_{2} = 
\frac{\sqrt{fh} r^{2} (f_{2,X}-h \phi'^{2} f_{2,XX})\phi' }{2}
+\frac{h^{3/2} r^{2} (6 f_{2,Y}+f_{2,X F}-6 h \phi'^{2} f_{2,X Y}) \phi' A_0'^{2}}{2 \sqrt{f}}
-\frac{Ph \phi' A_0'f_{2,X \tF}}{2}
\notag\\
&&\hspace{.75cm}
+\frac{h^{5/2} r^{2} (f_{2,F Y}-4 h \phi'^{2} f_{2,YY}) \phi' A_0'^{4}}{f^{\frac{3}{2}}}
-\frac{Ph^{2} \phi' A_0'^{3}f_{2,\tF Y}}{f}\,,\notag\\
&&
c_3=-\frac{r^2}{2}\sqrt{\frac{f}{h}}\frac{\D{\cal E}_{11}}{\D\phi}\,,\qquad 
c_4=0\,,\qquad 
c_5=-\left(\frac{1}{r}+\frac{f'}{2f}\right)a_4\,,\qquad
c_6=\frac{f+rf'}{2f}a_4-\frac{\phi'}{4}c_2+\frac{A_0'^2}{4}v_1+\frac{\phi'A_0'}{8}v_4\,,
\notag\\
&&
d_1=\frac{a_4}{2f}\,,\qquad 
d_2=0\,,\qquad 
d_3=\frac{2}{fhr\phi'}\left[(f'h-fh')a_4-\frac{P^2f(fhv_9+v_{10})}{r^3}\right]\,,\qquad
d_4=\frac{a_4}{r^2}+\frac{P^2}{r^4}v_{10}\,,
\notag\\
&&
e_1=\frac{r(f'h-fh')}{(fh\phi')^2}a_4\,,\qquad 
e_2=-\frac{1}{\phi'}\left(c_2+\frac{A_0'}{2}v_4\right)\,,\qquad
e_3=-\frac12\frac{\D{\cal E}_{\phi}}{\D\phi}\,,
\notag\\
&&
e_4=\frac{1}{fh^2r\phi'^2}\left[(fh'-f'h)a_4+\frac{(hr^4A_0'^2+P^2f)(fhv_9+v_{10})}{r^3}\right]\,,
\notag\\
&&
v_1=
\frac{\sqrt{h} \left[(f_{2,F}-2 h \phi'^{2} f_{2,Y}) r^{4}+P^{2} f_{2,\tF\tF}\right]}{2 \sqrt{f} r^{2}}
+\frac{h^{3/2} r^{2} (f_{2,FF}-4 h \phi'^{2} f_{2,F Y}+4 h^{2} \phi'^{4} f_{2,YY}) A_0'^{2}}{2 f^{3/2}}
\notag\\
&&\hspace{.75cm}
-\frac{Ph (f_{2,F \tF}-2 h \phi'^{2} f_{2,\tF Y}) A_0'}{f}\,,
\notag\\
&&
v_2=A_0'v_1\,,\qquad 
v_3=-A_0'v_1-\frac{\phi'}{2}v_4\,,
\notag\\
&&
v_4=
h\phi' \left[
P f_{2,X \tF}
-\frac{\sqrt{h} r^{2} (4 f_{2,Y}+f_{2,X F}-2 h \phi'^{2} f_{2,X Y}) A_0'}{\sqrt{f}}
-\frac{2 h^{3/2} r^{2} (f_{2,F Y}-2 h \phi'^{2} f_{2,YY}) A_0'^{3}}{f^{\frac{3}{2}}}
+\frac{2P h A_0'^{2}f_{2,\tF Y}}{f}
\right]\,,
\notag\\
&&
v_5=\sqrt{\frac{h}{f}}r^2A_0'(f_{2,\phi F}-2h\phi'^2f_{2,\phi Y})-Pf_{2,\phi \tF}\,,\qquad 
v_6=0\,,\qquad 
v_7=\frac{A_0'^2}{4}v_1\,,\qquad
v_8=\frac{2A_0'}{f}v_{10}\,,
\notag\\
&&
v_9=\frac{f_{2,F}}{2\sqrt{fh}}\,,\qquad
v_{10}=-\frac{\sqrt{fh}(f_{2,F}-2h\phi'^2f_{2,Y})}{2}\,,\qquad 
v_{11}=0\,,\qquad 
v_{12}=0\,,\qquad 
v_{13}=-\frac{2A_0'(fhv_9+v_{10})}{fh\phi'}\,,
\notag\\
&&
m_1=
\frac{P\sqrt{f} f_{2,F}}{2 r^{2} \sqrt{h}}
-\frac{\sqrt{h} P (f_{2,FF}-2 h \phi'^{2} f_{2,F Y}-f_{2,\tF\tF}) A_0'^{2}}{2 \sqrt{f} r^{2}}
-\frac{h (f_{2,F \tF}-2 h \phi'^{2} f_{2,\tF Y}) A_0'^{3}}{2 f}
+\frac{P^{2} A_0'f_{2,F \tF}}{2 r^{4}}\,,
\notag\\
&&
m_2=
-\frac{P^{2} A_0'f_{2,F \tF}}{2 r^{4}}
-\frac{P \left[f f_{2,F}+f \phi'^{2} h f_{2,X F}-h (f_{2,FF}-f_{2,\tF\tF}-4 h \phi'^{2} f_{2,F Y}) A_0'^{2}\right]}
{2 \sqrt{fh} r^{2}}
\notag\\
&&\hspace{.9cm}
-\frac{h A_0' \left[f\phi'^{2}f_{2,X \tF} -(f_{2,F \tF}-4 h \phi'^{2} f_{2,\tF Y}) A_0'^{2}\right]}{2 f}\,,
\notag\\
&&
m_3=\frac{2P}{fr^2}v_{10}\,,\qquad 
m_4=\frac{2}{A_0'}\left(\frac{Pfv_9}{r^2}-m_1\right)\,,\qquad 
m_5=(f_{2,\tF})'+m_4'\,,\qquad 
m_6=\frac{2P}{r^2}v_9\,,\qquad
m_7=-m_4\,,
\notag\\
&&
m_8=\frac{2P}{r^2}v_{10}\,,\qquad 
m_9=\frac{P\sqrt{h} \phi' (f f_{2,X F}+2 h A_0'^{2} f_{2,F Y})}{\sqrt{f} r^{2}}
+\frac{h \phi' A_0' (f f_{2,X \tF}+2 h A_0'^{2} f_{2,\tF Y})}{f}\,,
\notag\\
&&
m_{10}=\frac{Pf}{r^2A_0'}v_{13}\,,\qquad 
m_{11}=-\frac{P\sqrt{f}f_{2,\phi F}}{\sqrt{h} r^2}-A_0'f_{2,\phi\tF}\,,\qquad
m_{12}=\frac{P}{r^2}v_{13}\,.
\ea
%

\bibliographystyle{mybibstyle}
\bibliography{f2perturbation}

\end{document}